\documentclass[useAMS,usenatbib]{mn2e}
\usepackage{graphicx}

\def\ltsima{$\; \buildrel < \over \sim \;$}
\def\simlt{\lower.5ex\hbox{\simlt}}
\def\gtsima{$\; \buildrel > \over \sim \;$}
\def\simgt{\lower.5ex\hbox{\gtsima}}

\def\kpc{{\rm\,kpc}}
\def\msun{{\rm\,M_\odot}}

\title[Canis Major Red Clump stars]
{The core of the Canis Major galaxy as traced by Red Clump stars}
\author[M. Bellazzini et al.]{
M. Bellazzini$^{1}$\thanks{E-mail: michele.bellazzini@bo.astro.it}, R. Ibata$^{2}$,
N. Martin$^{2}$, G.F. Lewis$^{3}$, B. Conn$^{3}$, M.J. Irwin$^{4}$\\
$^{1}$
INAF - Osservatorio Astronomico di Bologna, via Ranzani 1, 40127, Bologna,
Italy\\
$^{2}$
Observatoire de Strasbourg, 11, rue de l'Universit\'e, F-67000, Strasbourg, 
France\\
$^{3}$
Institute of Astronomy, School of Physics, A29, University of Sydney, NSW
2006, Australia\\
$^{4}$
Institute of Astronomy, Madingley Road, Cambridge, CB3 0HA, U.K.\\
}
\date{Accepted for publication by MNRAS, 30 November 2005}

\begin{document}

\pagerange{} \pubyear{2005}

\maketitle

\label{firstpage}

\begin{abstract}
The  recently-discovered stellar  system  in Canis  Major is  analysed
using He-burning  Red Clump stars as  tracers. Canis Major turns out to be 
the strongest and most spatially confined overdensity of the whole Galactic
Disc, both in terms of number density and of statistical significance.
When projected onto the Galactic Plane, it appears as an elongated and compact
overdensity extending from $l \sim 200\degr$ to $l\sim 280\degr$ with a 
roundish core toward $l\sim 240\degr$.  
We find  that the main
body of  the system has an integrated  absolute magnitude $M_V=-14.4\pm
0.8$,  a  central surface  brightness  $\mu_{V,0}\simeq 24.0\pm  0.6$ 
mag/arcsec$^2$ and  a
line-of-sight profile peaked at  $D_{\sun}=7.2 \pm 1.0 \kpc$ with Half
Width at Half Maximum $\sim 2.0 \kpc$, in excellent agreement with the
results  obtained with  widely different  tracers (M  giants  and Main
Sequence stars) in  previous analyses.  The mean distance  to the main
body of Canis  Major is observed to increase  with increasing Galactic
longitude, from  $D_{\sun}\simeq 6.3  \kpc$ at $l\simeq  225\degr$, to
$D_{\sun}\simeq  9.3 \kpc$  at $l\simeq  265\degr$, in  good agreement
with the predictions of our  more recent N-body simulation that models
CMa as a  dwarf galaxy being accreted in a planar  orbit onto the disc
of the Milky Way.  We confirm  that the Canis Major system has all the
characteristics of  the relic  of a  dwarf galaxy seen  on top of a
large-scale overdensity that  we detect all over the third and fourth
Galactic quadrants ($180\degr  \le l\le 360\degr$, with a strong
maximum around $l=290\degr$ and  $b\ga -5\degr$) that is identified as
the stellar component of the  southern Galactic Warp.
On the other hand, the possibility that a peculiar deformation/asymmetry 
of the outer Galactic Disc may be at the origin of the observed 
distribution of overdensities toward CMa cannot be definitely ruled out 
with the data presented in this paper.
We also address a recent
claim  that Canis  Major  is on  the  outskirts of  a larger  ``Argo''
structure centred  at $l\simeq 290\degr$. Our analysis  shows that the
stellar populations  in the latter  are distributed over a  very large
distance range  along the  line of sight,  and do  not give rise  to a
narrow peak in density in this direction, contrary to what is observed
in Canis Major. This suggests that the Argo structure is likely due to
Galactic asymmetries such as the Warp.
\end{abstract}

\begin{keywords}
Galaxy: structure - galaxies: dwarf - open clusters: general - 
open clusters: individual: Tombaugh~2, AM-2, Heffner~11
\end{keywords}

\section{Introduction}

During the last decade, it has  been fully realised that the relics of
the latest  accretion events that  contributed to the assembly  of the
Milky  Way can  be observationally  identified and  studied  in detail
\cite[see,  for  example][and  references  therein]{ibata01,  newberg,
majewski}.   These  findings   simultaneously  provide  a  qualitative
confirmation and  a new challenge for the  current cosmological models
in which  the growth  of large  galaxies is driven  by the  process of
hierarchical  merger   of  sub-units  \citep{wr78,wf91}.    The  tidal
disruption of dwarf galaxies within the Galactic potential may lead to
the production of long-lived stellar  streams \citep[as in the case of
the  dwarf  spheroidal Sagittarius  galaxy  -  Sgr dSph,  see][]{il98,
ibata01, ibata02,  ivez, newberg,  majewski, bell03a} that  may reveal
fundamental  information about  the  process of  disruption, the  mass
distribution within  the Galactic halo  of Dark Matter, its  degree of
clumpiness, etc.  \cite[see][]{ibata01, helmi04, katysgr, law}.

The recent discovery  of a large stellar stream  nearly coplanar with
the Galactic disc \cite[the  Canis Major/Monoceros Ring, hereafter the
Ring, for brevity][]{newberg, yanny, ibaring, majewski, rochap, crane,
conn, connb}
suggests that  the accretion of small  stellar systems may  have had a
considerable r\^ole also in the assembly of the disc components of the
Milky Way (in  particular, the Thick Disc) as  envisaged by the latest
detailed models of disc-galaxy formation within a cosmological context
\citep{abadi, abadi2, helmi}.

In \citet[][hereafter Pap-I]{martin} we reported on the identification
of a possible new stellar relic (the Canis Major  dwarf galaxy, hereafter CMa)
located  at $7-8  \kpc$ from  the  Sun whose  approximate centre  lies
around  $(l;b)\sim  (240\degr  ;   -7\degr)$,  and  that  may  be  the
progenitor  of  the  Ring  \cite[see  also][]{delgado-mod,dana}.   CMa  was
identified as  a strong  elliptical-shaped overdensity of  M-giants by
the  comparison  of star  counts  in  Northern  and Southern  Galactic
hemispheres  \citep[from  the 2MASS  All  Sky Survey,][]{cutri}.   The
structure was  suggested to be possibly associated  with some globular
\cite[but see][]{delgado-mod,delgado-pm} and open clusters  
\citep{martin, moon, crane}. 
The relic  appears to have luminosity and  size similar to the
Sgr  dSph  (Pap-I).  The  optical  Color  Magnitude  Diagram (CMD)  we
obtained  in \citet[][hereafter  Pap-II]{moon} revealed  a  narrow and
well-defined Main Sequence typical of an intermediate to old (age$\sim
4-10$ Gyr)  and moderately metal deficient  ($[M/H]\sim -0.5$) stellar
system;  these results  have been  fully confirmed  by the  deeper CMD
presented by \citet{delgado-cmd}. A Blue Plume of possibly young stars
or blue stragglers  - also typical of dwarf  spheroidal galaxies - has
been  detected in  both optical  CMDs. \citet{carraro} detected the
Blue Plume population in the background of several open clusters in the
third Galactic quadrant. While these authors conclude that this population
should be associated to the Norma and Perseus spiral arms of the Milky
Way, it turns out that all the considered fields lie within the region
of the sky where CMa is detected (from their Table~1 
$219\degr \le l\le 254\degr$; 
$-6.3\degr \le b\le +1.8\degr$, see Pap-I and Sect.~3, below), and in the same 
range of distances as CMa (6.0 kpc $\le D_{\sun}\le$ 11.7 kpc, see Pap-I, Pap-II
and Sect.~4, below). Moreover, the Blue Plume populations studied by 
\citet{carraro} appear to follow a distance - Galactic longitude trend quite
similar to that found here for the Red Clump population of CMa (see Sect.~4.1.1,
below).

The  first results  of  a large
spectroscopic  survey  provided an  estimate  of  the systemic  radial
velocity and  velocity dispersion of CMa \cite[$V_r\simeq  110 $ km/s;
$\sigma= 13.0$ km/s][hereafter Pap-III]{martinb}, while an estimate of
the  systemic  proper  motion  has  been  obtained  by  \citet{moma4}.
\citet[][hereafter  Pap-IV]{martinc} analysed  the kinematics  of more
than 1400  Red Giant Branch and  Red Clump stars around  the centre of
CMa, confirming the essence of the results of Pap-III, namely that CMa
stars  have  a  systemic  velocity  and a  velocity  dispersion  quite
different  from  what  expected   from  Galactic  disc  stars  at  the
considered  distance and  quite  typical of  Galactic dwarf  satellite
galaxies.  Moreover  they detected a clear distance  - radial velocity
gradient  among CMa  stars  that can  be  explained as  the effect  of
on-going  tidal  disruption  of  the  stellar  system.  A  significant
refinement of the proper motion  estimate was also obtained in Pap-IV.
Finally, \citet{dana} very recently obtained a new accurate estimate
of the proper motion of CMa and concluded that (a) the system has a 
significant component of motion perpendicular to the Galactic Plane 
($W\simeq -50$ km/s) and that (b) this W component of the space motion of 
CMa is in the opposite direction with respect to the expected W motion of 
stars associated with the Galactic Warp.
Other  interesting  investigations  of  the CMa  relic,  not  strictly
related to  the present work or in a too preliminary stage to be fully
useful for the analysis, can be found in  \citet{kinman, forbes,
boniUVES,mateu}.

\citet{moma4} suggested that  the CMa overdensity may be  an effect of
the stellar component of  the Galactic Warp \citep{djorg, yusi, lopez,
vig},  whose   broad  southern  maximum  (as   a  South/North  density
asymmetry)  lies   in  the  range   $240\degr  \le  l   \le  300\degr$
\citep{lopez, vig}.   In Pap-II and Pap-III, in  particular, we showed
that  such  a   conclusion  can  be  reached  only   if  the  distance
distribution of the  adopted tracers (M giants) is  neglected and that
CMa  is in  fact an  additional overdensity  of different  nature with
respect to  the large scale  Warp of the  Galactic disc. On  the other
hand the  disentanglement of the  two structures may  prove difficult,
limiting our  possibility to obtain  a detailed global  description of
CMa  and  the Warp  (see  Pap-III,  for  this discussion).   A  global
analysis of CMa is also  hampered by the large degree of contamination
from Galactic  stars, by the  high and variable extinction  toward the
Galactic  plane and  by its  proximity  to us  that imply  a very  low
surface  density in  star  counts  per unit  solid-angle  of sky  (see
Sect.~4.4, below).

It is clear  that a large scale study of CMa  using a different tracer
than the M-giants used in the previous wide-field studies of the system 
would provide a very useful independent check
of the  results of Pap-I  and may help  us to understand  the relation
between CMa and the Warp.  In  Pap-I we showed that the Red Clump (RC)
of Helium-burning stars  of the CMa system is  clearly detected in the
2MASS database and in Pap-II we used  the RC stars to show that CMa is
an independent  structure, superposed on  the Galactic Warp.   Here we
extend the  use of RC  stars to trace  the whole structure of the outer
Galactic Disc, focusing on the CMa relic and  of the Galactic Warp in 
its surroundings.  In  Sect. 2 we
describe  the adopted dataset  and assumptions,  and we  introduce our
analysis method. In Sect. 3 we study the spatial distribution of overdensities
rising from asymmetries between the Southern and Northern Galactic hemisphere,
focusing on the relation between Canis Major and the Galactic Warp.
In Sect. 4  we derive some fundamental parameters of CMa as a stellar system,
and, finally, in Sect. 5 we summarise and discuss the main results
of these analyses.

\begin{figure}
\includegraphics[width=84mm]{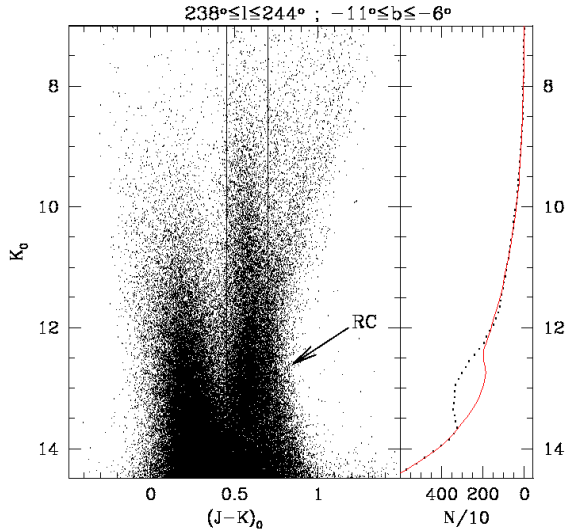} 
\caption{Left  panel: Infrared  Colour Magnitude  Diagram of  a $6\degr
\times 5\degr$ field  near the centre of the  CMa overdensity. The Red
Clump of CMa is indicated by the arrow. The vertical lines enclose the
colour  range in  which  we  select candidate  RC  stars. Right  Panel:
Luminosity Function  of the  selected RC stars (dotted black line).  
Note the  strong bump associated with the RC of CMa at $K\sim 13.0$.  
The Luminosity Function obtained exactely in the same way from the corresponding
synthetic sample extracted from the R03 model (see Appendix A) is plotted as a
red continuous line. The synthetic LF has been rescaled by a factor $1/3$ to
match the total number od stars in the two samples.}
\end{figure}

\section{Red Clump stars as tracers of outer Disc structures.}

We  extracted  from  the  2MASS  Point Source  Catalogue  the  sources
comprised between $l=40\degr$  and $l=320\degr$, having $-30\degr \le
b\le 30\degr$. The selected range of longitude and latitude cover the largest
part of the Galactic Disc, avoiding only the surroundings of the Galactic
Centre, where the stellar crowding and the interstellar extinction may prove too
severe for a useful analysis in the present context. 
Stars were selected according  to the quality of the photometry
and   to    avoid   spurious   and   contaminated    sources   as   in
\citet{bell03b}. Moreover, only stars having photometric uncertainties
$< 0.3$ mag both in J {\em and} K magnitude have been included in the sample.
The  magnitudes of all the stars were  corrected for
extinction using  the colour  excesses interpolated on  the COBE/DIRBE
maps \citep{schlegel} and modified  according to \citet{boni}, as done
in  \citet{moma4} and  Pap-III. Obviously a 2-dimensional extinction map
cannot account for the 3-dimensional distribution of the intersellar dust, thus
the adopted correction may not always be good for nearby stars, but should be
quite appropriate for the distant Red Clump stars we are interested in.  
The  extinction laws  by \citet{rl85}
have  been adopted. In  the following  J, H,  and K  magnitudes denote
reddening-corrected  magnitudes ${\rm  J_0}$, ${\rm  H_0}$,  and ${\rm
K_0}$, unless otherwise stated. Only stars with $K<15.0$ were retained for the
analysis.

In Figure~1 we show the J-K,  K CMD of a low-extinction field near the
centre of the CMa structure. The diagram is dominated by Main Sequence
(MS) stars of the Galactic disc for $J-K\la 0.5$, for $J-K\ga 0.8$ the
M  giants  are  contributed  by   both  the  disc  and  CMa,  and  the
intermediate colour  strip, $0.45\le J-K\le 0.70$, enclosed by continuous 
lines in Fig.~1, should be  dominated by RC stars  located at
various distances. While  the M giant sequence of  CMa can be revealed
only with the subtraction of a control field symmetric with respect to
the  Galactic Plane (Pap-I)  the Red  Clump of  the system  is clearly
visible  without subtraction  in the  range  $12\le K\le  14$ with  an
apparent peak around $K=13$, as  readily shown by the histogram in the
left panel of Fig.~1 (dotted black line), in  excellent agreement with Pap-I. 
Hence RC stars
are very promising  tracers of the CMa structure.  In the
following discussion we will  limit our analysis to the stars  in the colour strip
enclosing the typical RC stars, $0.45 \le J-K \le 0.75$, as in Pap-II
(e.g., the region enclosed within parallel continuous lines in Fig.~1). 
In the left panel of Fig.~1 we also plotted the
Luminosity Function (LF) obtained with the same selection criteria from 
a synthetic
sample of the same region of the sky extracted from the Galaxy model by
\citet[][hereafter R03, red continuous line; see Appendix A for details]{r03}. 
The R03 model includes a warped and flared disc
(whose parameters have been fixed according to observations) and the effect
of the displacement of the Sun with respect to the Galactic Plane by 15-16 pc
\citep[toward the Galactic North,][R03]{hamm}. 
It is interesting to note that the R03 model reproduces
excellently the observed LF for $K<12.3$ and $K>13.7$, but completely lacks the
strong bump around $K=13$ that we attribute to the RC of Canis Major.

With respect to M giants, RC  stars may suffer from a larger degree of
contamination by  unrelated sources. While  for $J-K\ge 1.0$  the only
possible contaminants  for M giants  are M dwarfs at  faint magnitudes
($K\ge 12.0$), our  colour-selected RC sample may be  contaminated by K
giants essentially  at any magnitude, and significant numbers of both M and G  
dwarfs can be pushed   into  the   RC  colour   range  by   photometric   
errors  for $K>14.0$ \cite[see][and Appendix A, for discussion]{lopez}. 
However  all the results  presented below  are in
the form of  subtractions of star counts in  the Southern and Northern
Galactic hemispheres.  Since the degree of contamination from Galactic
sources  as  a  function  of  magnitude  should  be  similar  in  both
hemispheres,  the  subtraction  should  minimise any  spurious  effect
associated with  contaminants (see Appendix A, below). 
Moreover, all the results presented below 
remain unaltered if we limit our sample to RC stars having $K\le
14.0$, where the contamination by dwarfs is not overwhelming \citep{lopez}.
In Appendix A we study the problem of contamination by dwarfs in some 
detail, showing that (a) in spite of the large fraction of dwarf that may
contaminate our samples and of the presence of North-South differences
in the degree of contamination, the adopted technique traces trustworthy real
asymmetries in the density of distant giants, for $K\le 14.0$, and (b) the
intrinsic asymmetry due to the $\simeq 15$ pc displacement of the Sun above the
Galactic plane \citep{hamm} has a negligible effect on our analysis.

On  the other hand  RC stars  should be
intrinsically  more  numerous  than   M  giants  and  they  provide  a
completely independent  and more  reliable distance scale  compared to
the   photometric   parallax  of   M   giants  (see   \citet{bell03b},
\citet{majewski}, Pap-I, Pap-II and Pap-III).

If not otherwise stated, we will  use in the following analysis only stars with
$|b|>5.0\degr$  to avoid  the  regions where  the  extinction is  most
severe. Fig. 2 shows that once the $|b|>5.0\degr$ strip is excluded, the
region considered in the present analysis is largely free from high extinction
regions (e.g. where $A_K>0.3$ mag). Since most of the structures we are
interested in are enclosed in the latitude strip $|b|\le 15.0\degr$, most of the
following analysis is limited to this range. 
Note that the  2MASS Point Source
Catalogue  is complete  down  to $K_S=14.3$  \citep{cutri}, hence  the
luminosity range in which most of the RC stars we will consider 
lie ( $11.0\la K\la 14.0$) is not expected to suffer from 
problems of incompleteness.

\begin{figure}
\includegraphics[width=84mm]{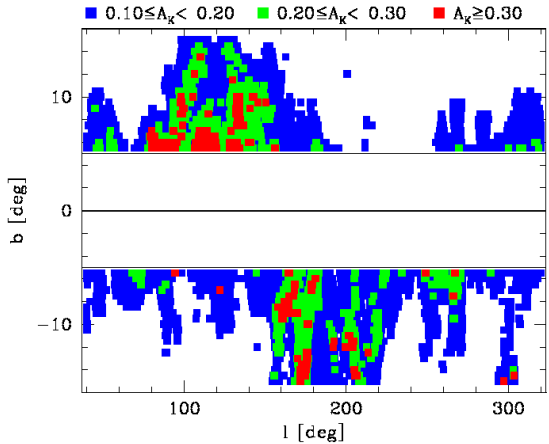} 
\caption{Map of the extinction (in K band) for the range of Galactic longitude
and latitude considered in the present analysis. The COBE-DIRBE reddening maps
\citep{schlegel} have been interpolated on a grid whose knots are spaced by
0.5 deg both in $l$ and $b$.}
\end{figure}

\subsection{CMa and the Warp as South-North overdensities}

It has to be noted that  the origin of disc warps and the relationship
between   gaseous   and  stellar   warps   is   far  from   understood
\citep{binney,  kuij}.  Moreover  the characterisation  of  the stellar
Warp of the  Milky Way is quite poor and  uncertain, mainly because of
the  problems inherent  with the  analysis  of stellar  fields at  low
Galactic  latitudes. Finally,  it  is important  to  realise that  the
adopted parametrisation may  describe different characteristics of the
warp and may induce some confusion  in the use of terms. For instance,
\citet{djorg} identified the stellar warp as a sinusoidal variation of
the mean latitude of their tracers (namely IRAS sources) as a function
of longitude.  In  this case the "maximum of  the warp" coincides with
the maximum  {\em angular height} above the  Galactic equator, similar
to the parametrisation adopted when  dealing with the gaseous warp. On
the  other hand, \citet{lopez}  described the  warp as  the sinusoidal
variation  of  the ratio  of  star  counts  in Northern  and  Southern
Galactic hemisphere  as a function of longitude,  hence their "maximum
of  the  warp"  is  the   region  where  the  Southern  (or  Northern)
projected overdensity reaches its maximum.  
The two definitions may provide very
different views,  depending on  the structure of  the disc and  on the
relative 3-D  positions of the  observed structures and the  Sun. Since
the CMa galaxy  was identified as a South/North  overdensity it should
be  compared with  the  Warp as  a  South/North overdensity  \cite[\`a
la][]{lopez},  since  this is  the  only  relevant  parameter in  this
context.   The parametrisation  of the  amplitude of  the Warp  as the
artificial  shift to apply  to the  Galactic plane  to match  the star
counts  in   the  Northern  and  Southern   hemisphere  \cite[as  done
by][]{moma4} is not a fair description of South/North overdensities in
the present context,  since it turns into a  comparison of stars lying
at different distances, as shown  in Pap-III.  Hence, in the following
we  will   always  deal   with  CMa  and   the  Warp   as  South/North
overdensities.

\subsection{The construction of subtracted density maps}

Most of the following analysis will make use of subtracted density maps
obtained form the above described sample of color-selected RC stars. In this
section we will describe in detail how these maps are obtained.

RC  stars  are  widely  used  as distance  indicators  \cite[see,  for
example][and references therein]{pac, ata, babu, lopez}. In Pap-II we assumed
$M_K=-1.5  \pm  0.2$  for the  RC  of  CMa,  since, according  to  the
theoretical  models  by   \citet{sg02},  this  interval  encloses  the
absolute K  magnitude of the  RC for populations of  metallicity $-0.4
\la[M/H]\la -0.7$  and age between 4  Gyr and 10 Gyr,  i.e. the proper
age  and metallicity for  CMa (see  Pap-II). Here  we adopt  the same
assumption  to obtain the distance of each RC star in our sample.
These individual distances should
be  regarded with  some caution  since the  adopted distance  scale is
optimised for the RC of CMa and may provide misleading answers when is
applied (a) to contaminant sources that are not RC stars (see sect. 2,
above), and  (b) to  RC stars outside  the age and  metallicity ranges
outlined  above. On the other hand, the range of age and metallicity of stellar
populations in the outer Galactic Disc that produce RC stars should not be too
different from that outlined above. Hence, the assumed global distance scale
is expected to provide a reasonable description of the overall spatial 
distribution of RC stars in this region  of the Galaxy and a reliable 
representation of the CMa system. In particular, it should be considered that RC
stars are much  less  sensitive  to   metallicity  and  age  variations  in  the
considered populations compared to M-giants \cite[see,][and references
therein]{majewski,crane,martin,argus}. Finally, it should also be recalled that
in the following we will deal only with {\em subtracted} density distributions:
the subtraction should cancel out many of the above effects if, as it very
reasonable to assume, at a given Galactocentric distance the age and 
metallicity distributions are similar in both Galactic hemispheres.

We used  the
distances  of  individual RC  stars  in  our  sample to  obtain  their
Cartesian Heliocentric coordinates:

\begin{equation}
x=D_{\sun}cos(l)cos(b) [\kpc]
\end{equation}
\begin{equation}
y=D_{\sun}sin(l)cos(b) [\kpc]
\end{equation}
\begin{equation}
z=D_{\sun}sin(b) [\kpc]
\end{equation}

In this  system the coordinates  of the Sun  are (x,y,z)=(0.0,0.0,0.0) kpc
and  the centre of  the Galaxy  is located  at (x,y,z)=(+8.0,0.0,0.0) kpc,
having assumed  that the distance  to the Galactic center  is $8.0\kpc$.
To avoid any possible effect of local inhomogeneities in
the vicinity  of the Galactic Plane  we also removed  the stars having
$|z|<0.5$  kpc.  Given the  adopted selections  this is  equivalent to
excluding stars within $\simeq 1$ kpc from the Sun.

From these samples we computed the  density of RC stars on an x,y grid
with boxes of size $1\kpc \times 1\kpc$, spaced by $0.5\kpc$ both in x
and in y. The overlapping between adjacent boxes ensures some degree of
smoothing, providing more clearly readable maps.
In this way we obtained x,y density maps of the Southern and
Northern  Hemisphere  that  can  be subtracted,  "pixel-to-pixel",  to
obtain a  map of  the residual  surface density in  the x,y  plane, that is
residual maps of the Galactic Plane as seen from the Galactic pole. 
The density at each point of the subtracted map is computed as
$$\rho_{(x,y)}=(n^{South}_{(x,y)}-n^{North}_{(x,y)})$$
Note that $n^{South}_{(x,y)}$ and $n^{North}_{(x,y)})$ are a sort of 
``column'' density, since at the position ($x_0$;$y_0$) the
derived density is equal to the number of RC stars having 
$x_0-0.5\le x\le x_0+0.5$ and $y_0-0.5\le y\le y_0+0.5$, {\em at any} z within
the limits of our sample. 
The imposed selection in latitude ($5.0\degr < |b|\le 15.0\degr$) 
imply that the range in z sampled by different pixels of the grid
varies with their heliocentric distance. In particular, at $D_{\sun}=5.0$ kpc
the sampled range of z is 0.5 kpc \ltsima $|z|$\ltsima  1.5 kpc, at 
$D_{\sun}=10.0$ kpc: 0.9 kpc \ltsima $|z|$\ltsima  2.6 kpc, at 
$D_{\sun}=13.0$ kpc: 1.1 kpc \ltsima $|z|$\ltsima 3.3 kpc. 
It is important to realize that this has no
consequence for the subtracted densities $\rho_{(x,y)}$, since each pair of
pixels to be subtracted have exactly {\em the same volume}.

Since a large absolute density residual (e.g. in number) does not necessarily
imply a statistically significant signal, it is mandatory to obtain the 
subtracted density maps also in terms of Signal-to-Noise ratio, that is in units
of $\sigma$:

$$\rho_{(x,y)}=(n^{South}_{(x,y)}-n^{North}_{(x,y)})/\sigma_{(x,y)}$$ 
where
$$\sigma_{(x,y)}=\sqrt{n^{South}_{(x,y)}+n^{North}_{(x,y)}}~~.$$

Subtracted maps in terms of number density provide a physical idea of the
extent of the detected overdensities, while subtracted maps in $\sigma$ units
allow us to discriminate structures that are statistically reliable and 
significant from those that can arise from mere statistical fluctuations.
 
\section{Overdensities in the Outer Disc}

Fig.~3 shows the maps of residuals obtained by subtracting the Northern
hemisphere density map from the Southern hemisphere density map, in terms of
number density (panel a) and in units of $\sigma$ (panel b). We recall here that
the regions $0\degr \le l\le 40\degr$ and $320\degr \le l\le 360\degr$ are
excluded from the present analysis. In Appendix A it is shown that while
the effect of contamination by dwarfs should not significantly affect subtracted
density maps for $K\le 14.0$, it may become quite relevant at fainter
magnitudes. In this section we present subtracted maps that include also stars
having $14.0< K< 15.0$ to avoid abrupt distance thresholds in the proximity of
the regions of interest. We caution the reader that the reliability of the
maps may be drastically reduced beyond the heliocentric circle
corresponding to the $K=14.0$ limit that is always clearly shown in Figs.~3,
4 and 5 below, as a reference. According to the results of Appendix A, the 
contamination by dwarfs can seriously affect the maps also around $l=180\degr$.
On the other hand, the main structures we deal with in the following analysis
lie safely in the range of latitudes and distances that should be correctly
probed by our subtracted-density technique (see Appendix A).

\begin{figure}
\includegraphics[width=84mm]{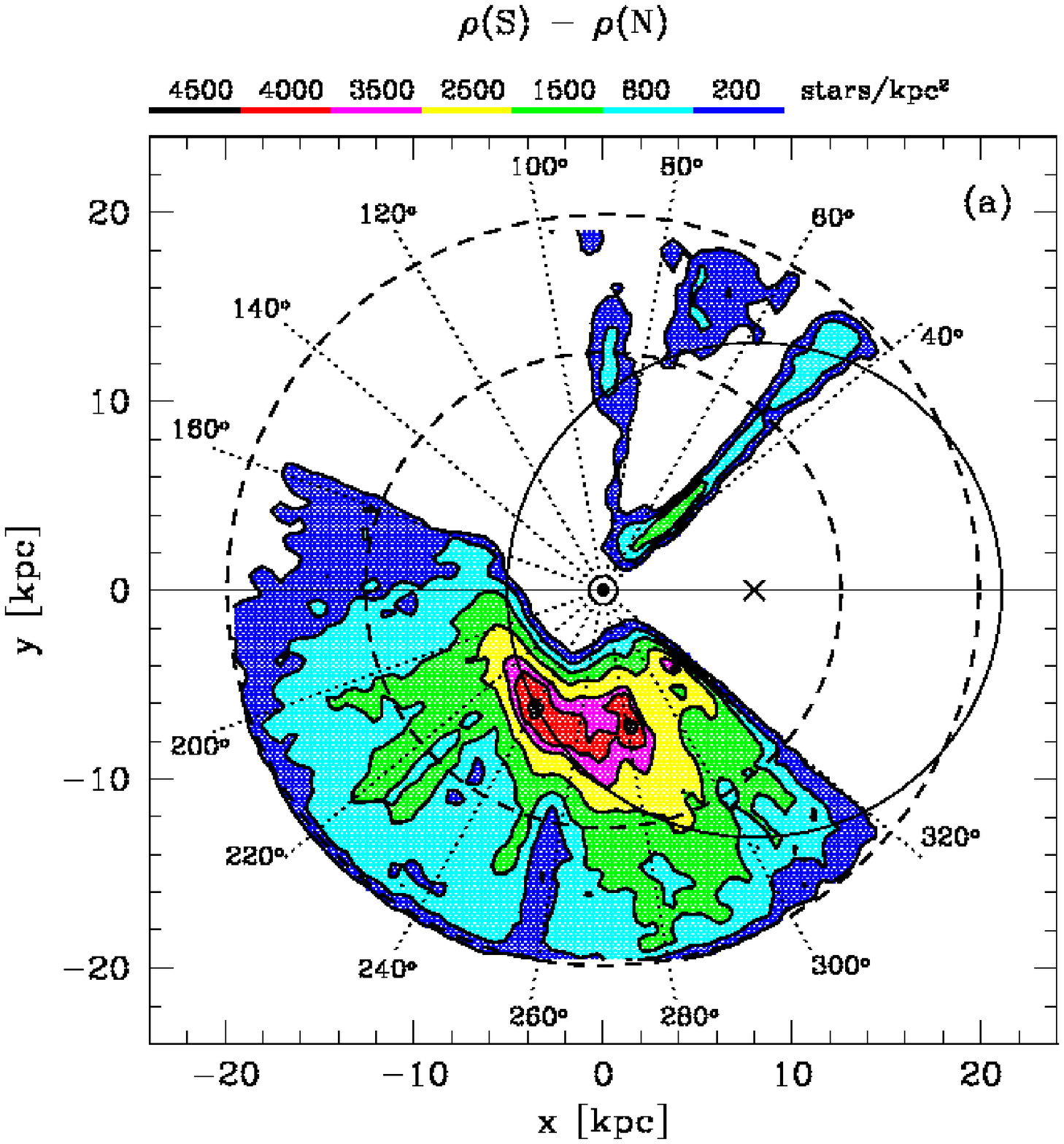} 
\includegraphics[width=84mm]{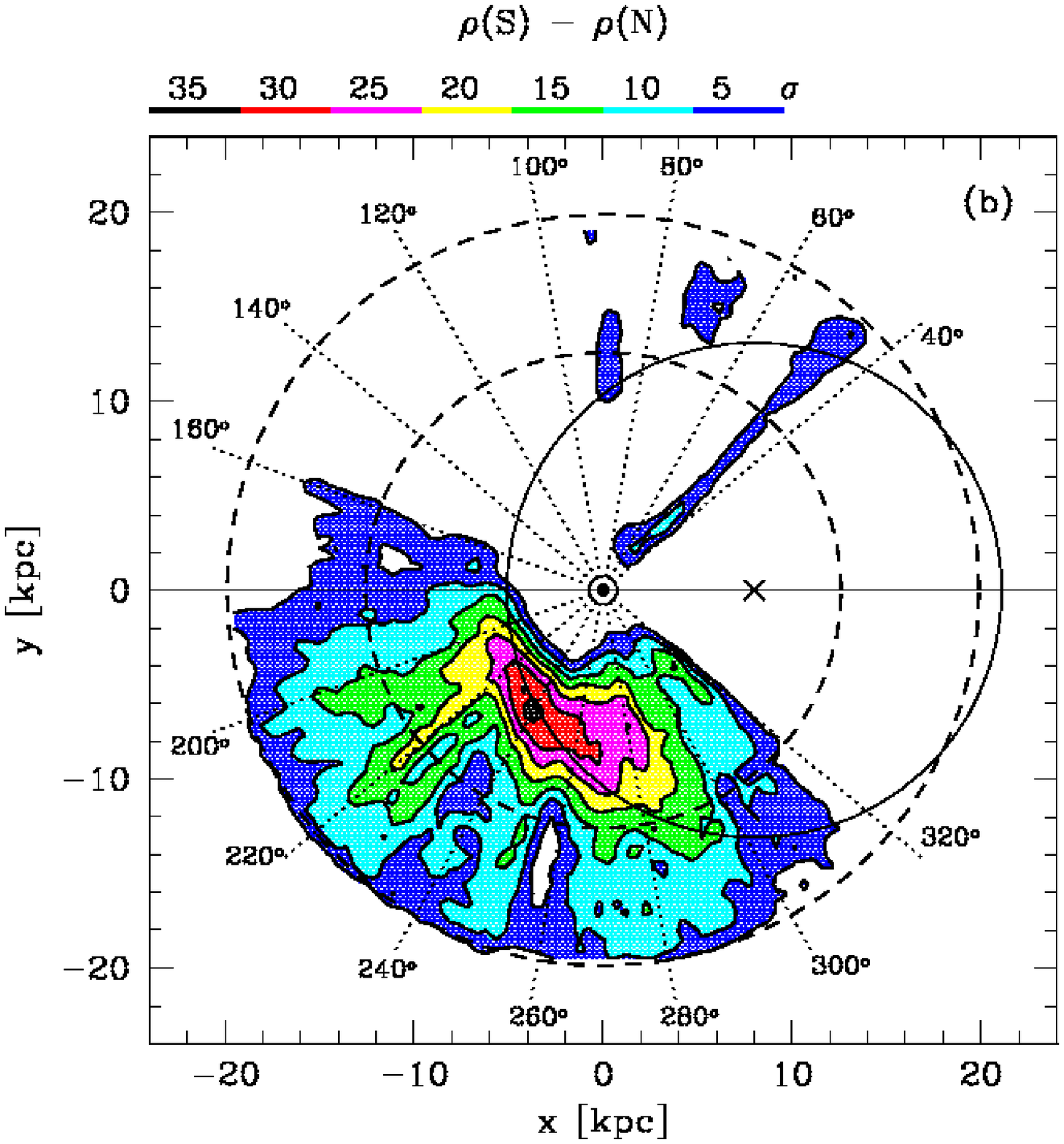} 
\caption{Subtracted density maps (South-North) in terms of number density (panel
a) and in units of $\sigma$ (Signal-to-Noise ratio; panel b). The direction of
Galactic longitudes from $l=40\degr$ to $l=320\degr$, with steps of $20\degr$,
are indicated as labelled dotted lines. The encircled dot marks the position of
the Sun, the $\times$ symbol indicates the Galactic Centre. Long-dashed circles
are centred on the Sun: the outer circle indicates the distance limit of 
our analysis, corresponding to RC stars having $K=15.0$, the inner circle
encloses the region sampled by RC stars having $K\le14.0$, i.e. the range in
which the contamination by dwarfs should be moderate. The continuous circle is
centred on the Galactic Centre and has a radius $R_{GC}=13.1$ kpc,
corresponding to the heliocentric distance of $D_{\sun}=7.2$ kpc for the centre
of the Canis Major structure at $l=240\degr$ (Pap III).
The region within $\pm 40\degr$ in longitude from the Galactic Centre is
excluded from the present analysis.}
\end{figure}

With the only exception of a few limited low-significance structures near the
edges of the map (in particular near $l=40\degr$, probably due to edge effects),
the strongest South-North overdensities are confined in the third and fourth
Galactic quadrants ($180\degr \le l\le 360$). This is not unexpected since it is
well known that the Galactic Disc is warped toward the South Galactic hemisphere
for $180\degr \le l\le 360$ and toward the North Galactic hemisphere for
$0\degr \le l\le 180$ \cite[][and references therein]{djorg,yusi,lopez,vig}.
The third and fourth Galactic quadrants appear dominated by a wide, 
galactic-scale overdensity that extends over the whole half of the disc with
$l>180\degr$. On top of this large-scale asymmetry there is a strong  and  
spatially confined  overdensity, with an elongated, approximately elliptical 
shape, with a denser roundish core located at  
(x,y)=(-3.7 kpc, -6.4 kpc), toward $l=240\degr$, at $R_{GC}\sim 13.0$
kpc and $D_{\sun}\sim 7.5$ kpc. This is the Canis
Major structure, as described in Pap-I, Pap-II and Pap-III. 
Note that a secondary core that appears towards $l\simeq 280\degr$ in Fig~3a is
revealed as a non-statistically-significant structure by the inspection of the
map in $\sigma$ units (Fig.~3b). 

The linear ``finger of God'' structures between
$l\sim 200\degr$ and $l\sim 220\degr$, and at $l\sim 250\degr$ correspond to
asymmetric features in the distribution of the extinction (see Fig.~2) and are
likely artefacts associated with the (unknown) three dimensional distribution of
the interstellar dust. These features are present in all the South-North maps
presented in this paper and will not be discussed further.
Note also that the strongest and most significant parts of the detected
overdensities (including the body of CMa) lie within the circle enclosing the RC
stars with $K\le 14.0$ (inner heliocentric long-dashed circle, at
$D_{\sun}=12.6$ kpc), that is in the
region where the contamination by dwarf stars is less severe
\cite[i.e., the contribution of dwarfs is less than 50 per cent, according to]
[and the contribution of dwarfs to subtracted densities in lower than 30
per cent, see Appendix A]{lopez}.

\begin{figure}
\includegraphics[width=84mm]{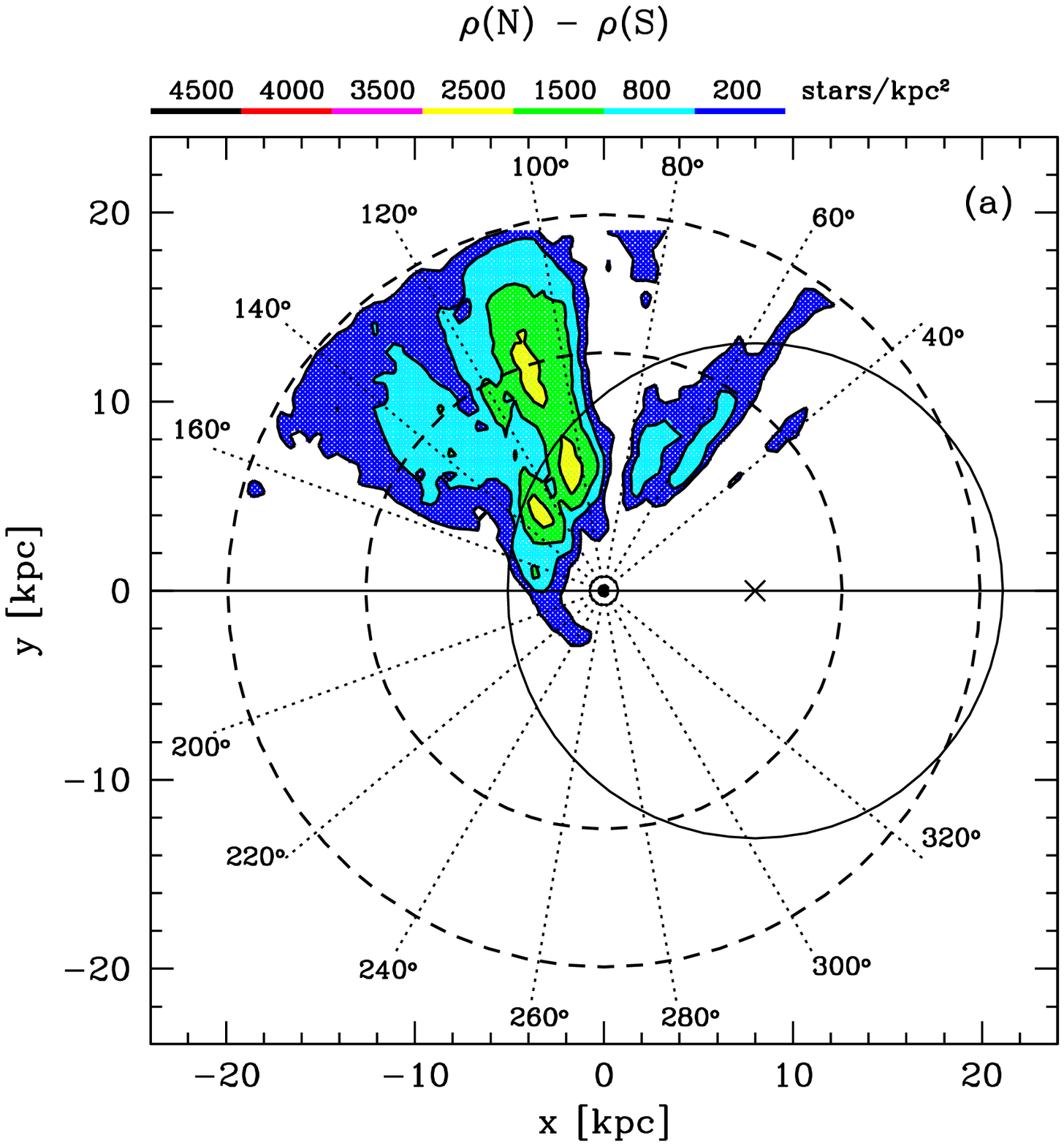} 
\includegraphics[width=84mm]{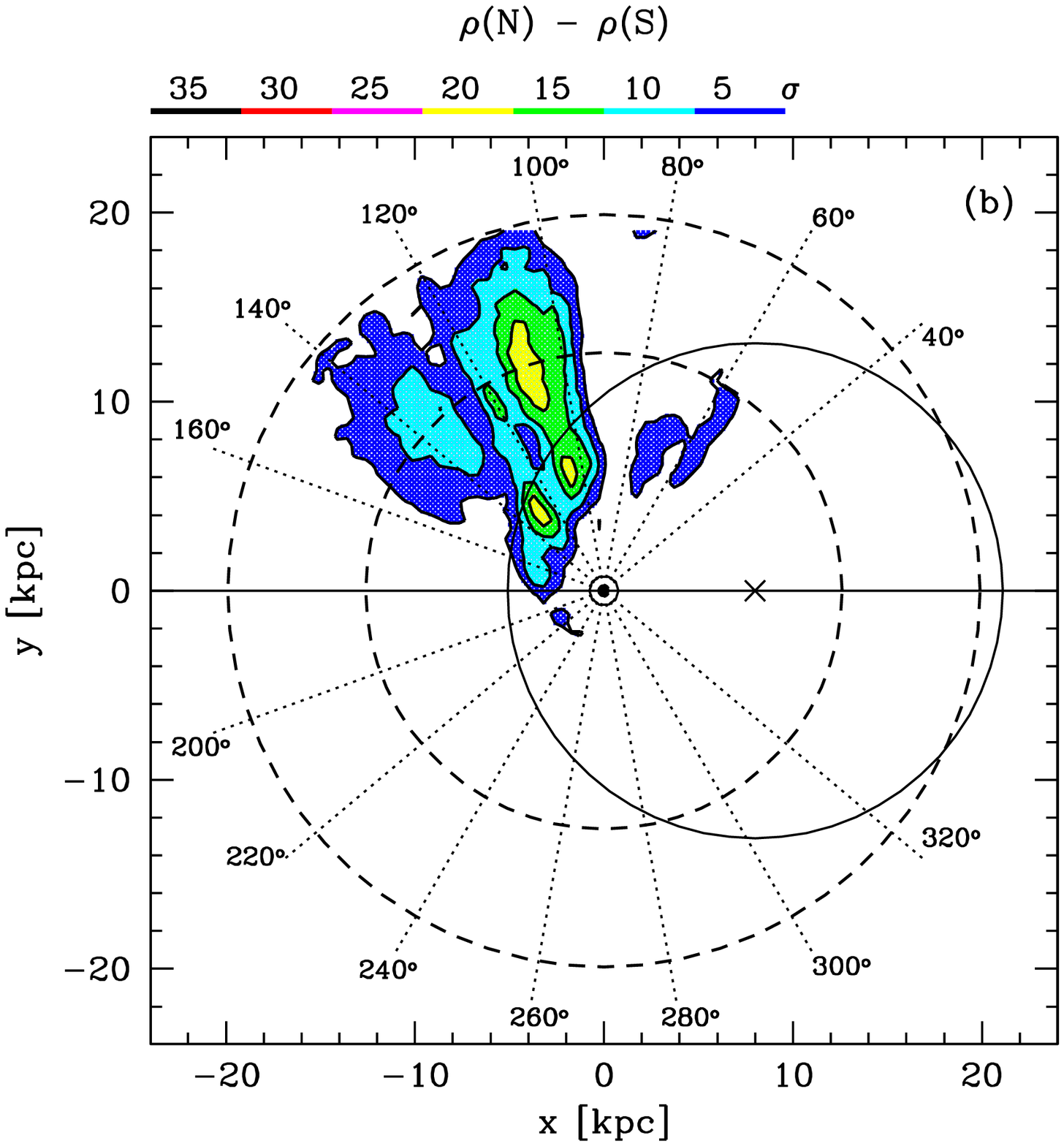} 
\caption{Subtracted density maps (North-South) in terms of number density 
(panel a) and in units of $\sigma$ (Signal-to-Noise ratio; panel b).
The symbols are the same as in Fig.~3.}
\end{figure}

Fig.~4 shows the maps of residuals obtained by subtracting the Southern
hemisphere density map from the Northern hemisphere density map, in terms of
number density (panel a) and in units of $\sigma$ (panel b). Also in this case
the detected overdensities are confined to the Galactic quadrants that are
expected to host the northern lobe of the Warp. The observed asymmetry has a
large scale - as in the South-North maps - but is weaker and less
significant than the large-scale Southern asymmetry. This may be due to
differences in the structure of the Northern and the Southern Warp 
\cite[see, for instance][]{yusi,porcel} and to the effect of the 
displacement of the Sun with respect to the Galactic Plane \citep{hamm}). 
However, in the present context is interesting to note that
(a) there is no overdensity comparable to CMa in the Northern Hemisphere and (b)
the most significant asymmetries lie in the range $100\degr\le l\le 140\degr$,
that is at $\sim 180\degr$ from the densest part of the large-scale Southern
overdensity ($280\degr \le l\le 320\degr$). 

The inspection of Fig.~3 and Fig.~4 clearly demonstrates that CMa is the
strongest (35 $\sigma$) overdensity of the whole Galactic Disc. Moreover it is
strongly spatially confined (the two innermost density contours have a typical
scale of $\sim$ 4 kpc $\times$ 2 kpc) and there is no corresponding structure
either on the opposite side of the disc ($l\sim 60\degr$) or at the reflected
position about the y or x axis.

Note that the detected excess of RC stars is quantitatively very
similar to what was found for M giants in Pap-I (in terms of $\sigma$ and of
fractional excess). CMa appears as a strongly concentrated, cored overdensity on
top of a galactic-scale asymmetry that has all the characteristics of the
Galactic Warp. 



\subsection{Removing the large scale asymmetry}

\begin{figure}
\includegraphics[width=84mm]{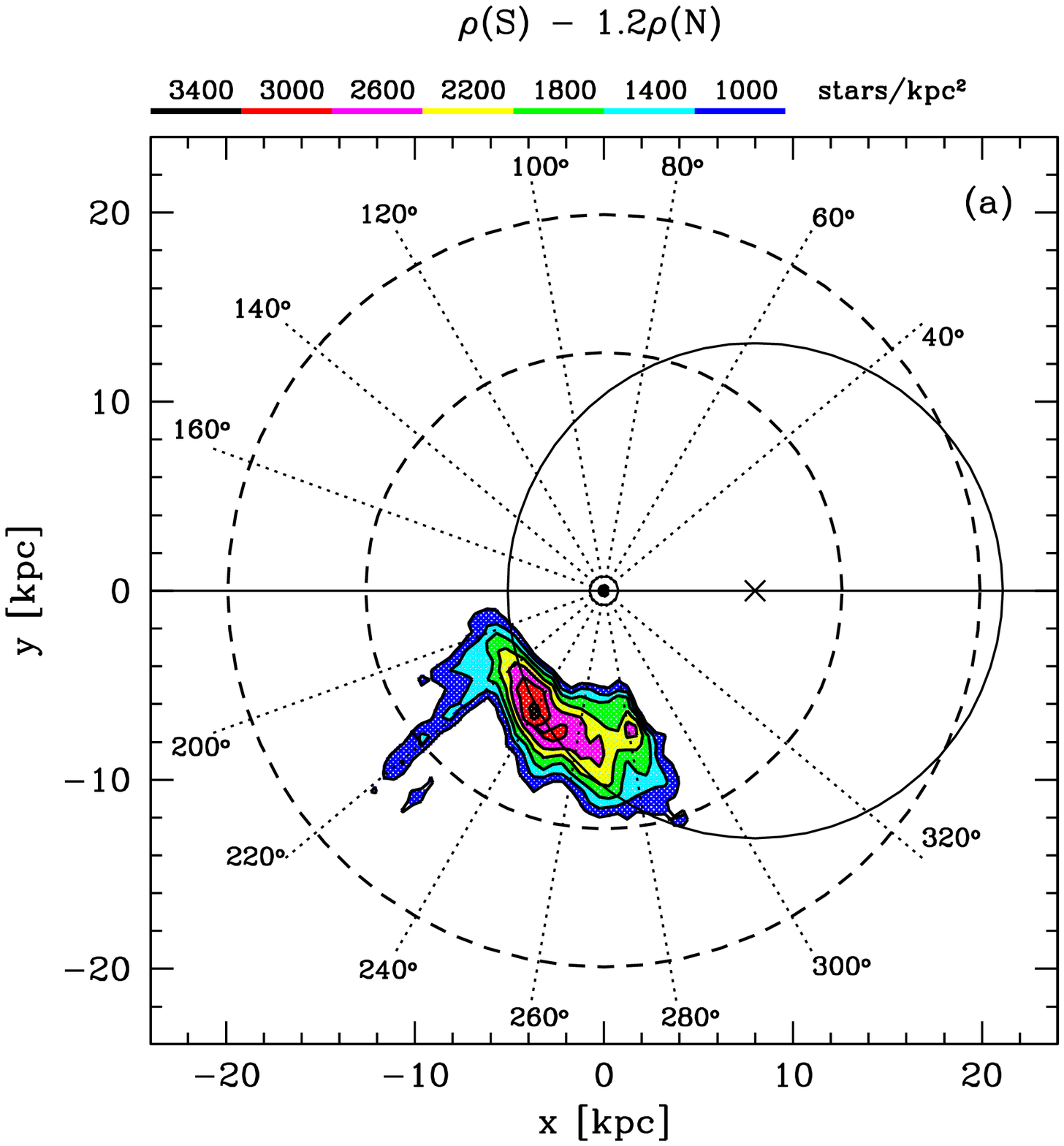} 
\includegraphics[width=84mm]{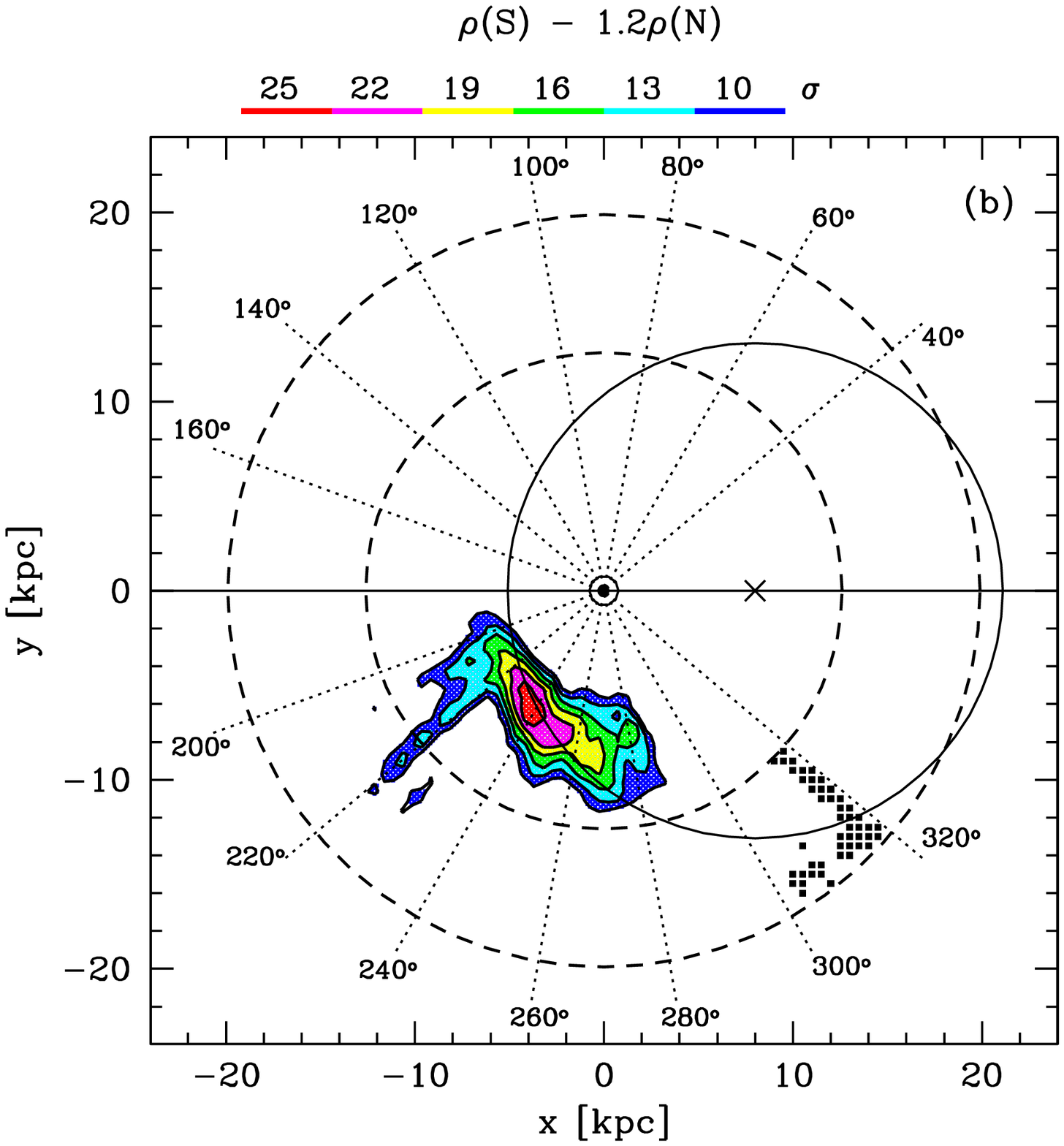} 
\caption{Subtracted density maps (South-North) in terms of number density 
(panel a) and in units of $\sigma$ (panel b). In this
case the densities in the Northern Hemisphere maps have been rescaled by a
factor $\times 1.2$, to attempt to remove the large scale asymmetry that we
identify as due to the Galactic Warp. In panel (b) we reported as small filled
squares the pixels of the map where {\em negative} residuals 
larger than $10\sigma$ are detected. All the other symbols are the same as 
in Fig.~3.}
\end{figure}

If the  global smooth overdensity is  related to the  Galactic Disc (i.e.
it reflects the Warp asymmetry) it
should be possible  to subtract it from the maps of Fig.~3. Since the
overall number  of southern RC stars  in our sample is  1.2 times that
found  in  the Northern  hemisphere,  as  the  simplest assumption  we
rescale the density of the Northern  map by this factor and repeat the
subtraction. The adopted normalization should account also for the effect
of the off-plane position of the Sun. 
The resulting  maps  of residuals  are  shown in  Fig.~5, both in
terms of number density (panel a) and in $\sigma$ units (panel b).
Adopting a safe  10$\sigma$ cut, the only surviving  structure is CMa
plus   tiny  traces   of   the  already discussed finger-of-God features at
$l\sim 220\degr$. 

Note that large negative residuals ($\ge 10\sigma$) appear only near the edges
of the  maps (Fig.~5b). 
This means  that the rescaled  northern density map  is a
reasonable  model  of  the  large-scale overdensity  observed  in  the
southern  hemisphere. We  tried  different simple  models fitting  the
observed overall decline of the large-scale overdensity with $R_{GC}$: 
in all cases the only overdensity
surviving the subtraction of the  model from the southern density map is
CMa,  in  all  cases  displaying  the characteristic  shape  shown  in
Fig.~5a,b.  
This  exercise strongly supports our interpretation of
the southern overdensities  in the considered region of  the Galaxy in
terms of  a large-scale Galactic asymmetry  (warp) plus  an additional
unexpected elongated compact structure with a dense core, i.e. the 
Canis Major overdensity.

\subsection{Comparison with a Warp model}

In response to the suggestion that the CMa overdensity may be associated with
the Galactic Warp, we have compared the observed characteristics of CMa with two
different models of the Galactic Warp, namely that enclosed in the Galactic
model by \citet{r03}, in Pap-II, and the warp model by \citet{yusi}, in Pap-III.
In both cases we found that the models were clearly unable to reproduce the
very compact structure of CMa. In particular the distance profile of the
overdensities predicted by models are much more extended than what is observed for
Canis Major. Here we extend the comparison to the model by 
\citet[][hereafter L02]{lopez} that has been constructed to fit the overall
distribution of RC stars in the Galactic Disc. 

\begin{figure*}
\includegraphics[width=168mm]{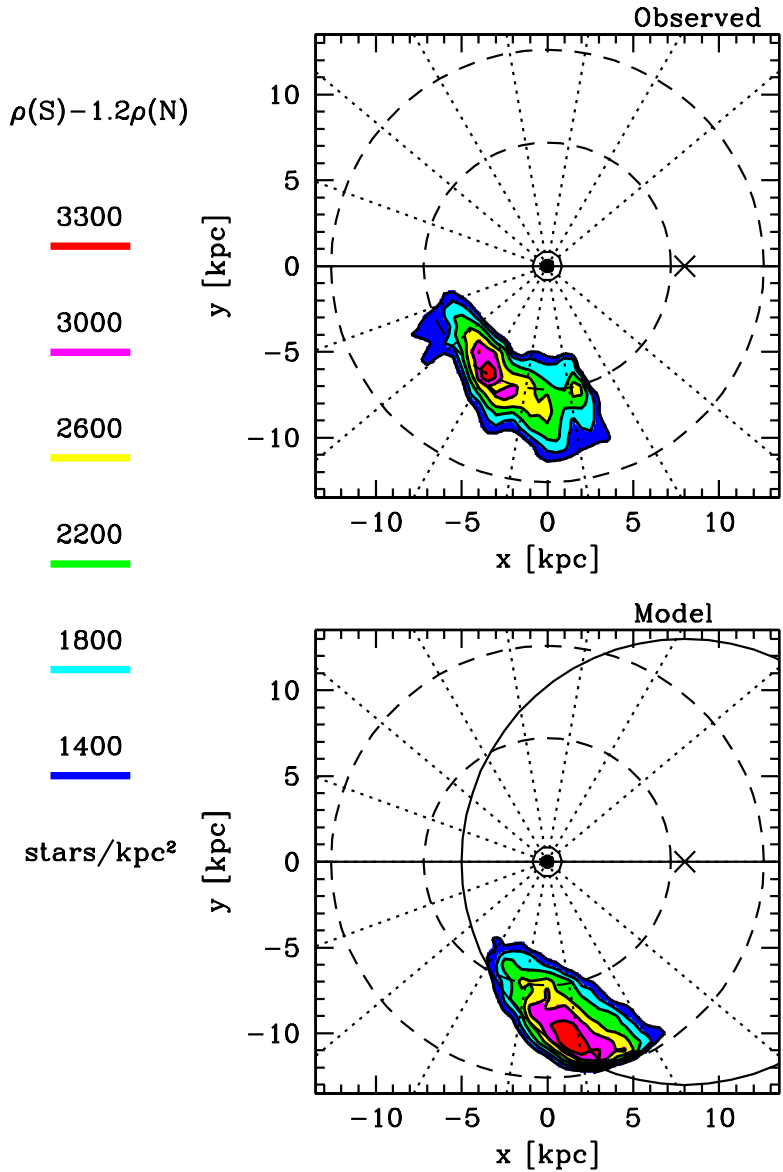} 
\caption{Subtracted density maps (South-North) 
from the observed sample (upper panel) and from the \citet{lopez} model (lower
panel). The Model map has been normalized to have the same maximum density as
the observed one.
The inner long-dashed circle have a radius $r=D_{\sun}=7.2$kpc, i.e. the
heliocentric distance of CMa (Pap-III). The outer long-dashed circles encloses
the region populated by RC stars having $K\le 14.0$. Both samples have also been
limited to $R_{GC}<17.0$ kpc. The continuous-line
galactocentric circle plotted in the lower panel of the figure, having radius
$=13.0$ kpc, shows that the relevant comparison is performed inside the
extrapolation region, hence the adopted form of the extrapolation cannot affect
our results.
Black dots in the lower panel
indicates negative residuals larger than 50\%, corresponding to the northern
lobe of the Warp.}
\end{figure*}

In order to compare density distributions obtained with the same technique, we 
constructed a Montecarlo realisation of the L02 model and we applied the same
selections as to the observed sample (i.e., $5\degr \le |b|\le 15\degr$ and
$|z|\ge 0.5$ kpc). We extended the L02 model outside its maximum
Galactocentric distance ($R_{GC}=13.0$ kpc) with the following assumptions for
the average elevation of the disc above the plane $z_w$ (see Eq.~20, in L02):

$$z_w=C_W R[kpc]^{5.25}sin(\Phi-\Phi_W)+15.0$$

for 13.0 kpc $< R_{GC}\le$ 16.0 kpc,\\ 
where $C_W=0.0012$ and $\Phi_W=-5.0$ deg, and

$$z_w=0.0$$

for $R_{GC}>16.0$ kpc. This extrapolation produces a regime of nearly constant
elevation for 13.0 kpc $< R_{GC}\le$ 16.0 kpc and a drop to the plane at larger
galactocentric distances, that appear as the most appropriate for the Southern
warp, according \citet{porcel} and \citet{yusi}, for example.
For homogeneity, we applied the $R_{GC}=17.0$ kpc cut also to the observed 
sample. Note however that the region relevant for the comparison is largely
included within the $R_{GC}=13.0$ kpc limits, hence the adopted extension of the
L02 model must have little effect on our results (see Fig.~6, below).

Finally, we limit the comparison to stars within $D_{\sun}=12.6$ kpc,
corresponding to the $K\le 14.0$ for RC stars.
After applying all of the above selection criteria, the Montecarlo realisation of the L02
model contains $10^6$ particles. Considering these particles as the analogues of
RC stars, we have obtained subtracted density maps from the model in exactly the
same way as we did for real stars (see Sect.~2.2). 
We normalized the density
scale of the Model map by requiring that the maximum density is the same in the
syntethic and observed map. We have tried also several other normalizations and
we found that the essence of the results remains unchanged.

In Fig.~6 the observed South-North subtracted map (upper panel) is compared to
the same map obtained from the L02 model. Both maps are represented in the same
scale of density excess. 
The differences between the two maps
are clearly evident. In the model map the Warp appears as a large wedge-shaped
overdensity, somewhat similar -- in shape -- to the densest part of the large scale
overdensity observed in Fig.~3. The density of this feature  
reaches a broad maximum in the range $270\degr \le
l \le 280\degr$, over a large range of Heliocentric distances, around 
(x,y)$_{[kpc]}\simeq$ (1.5;-10.3) -- while the maximum of the observed density
occurs at $l\simeq 240\degr$ and (x,y)$_{[kpc]}\simeq$ (-3.5;-6.2), more than 6
kpc apart. 
It is quite clear
that the model is totally unable to reproduce the CMa structure presented in the
upper panel of Fig.~6 as the only significant overdensity, 
extending from $l\sim 200\degr$ to $l\sim 275\degr$, with
its compact elliptical shape and short minor axis, and with its core at
$l=240\degr$. The differences between the two
structures (CMa and the L02 Warp model) clearly suggests that 
they are of a different nature and cannot be identified with one another, 
confirming the results presented in Pap-II and Pap-III (see also Fig.~1, here).

\begin{figure*}
\includegraphics[width=168mm]{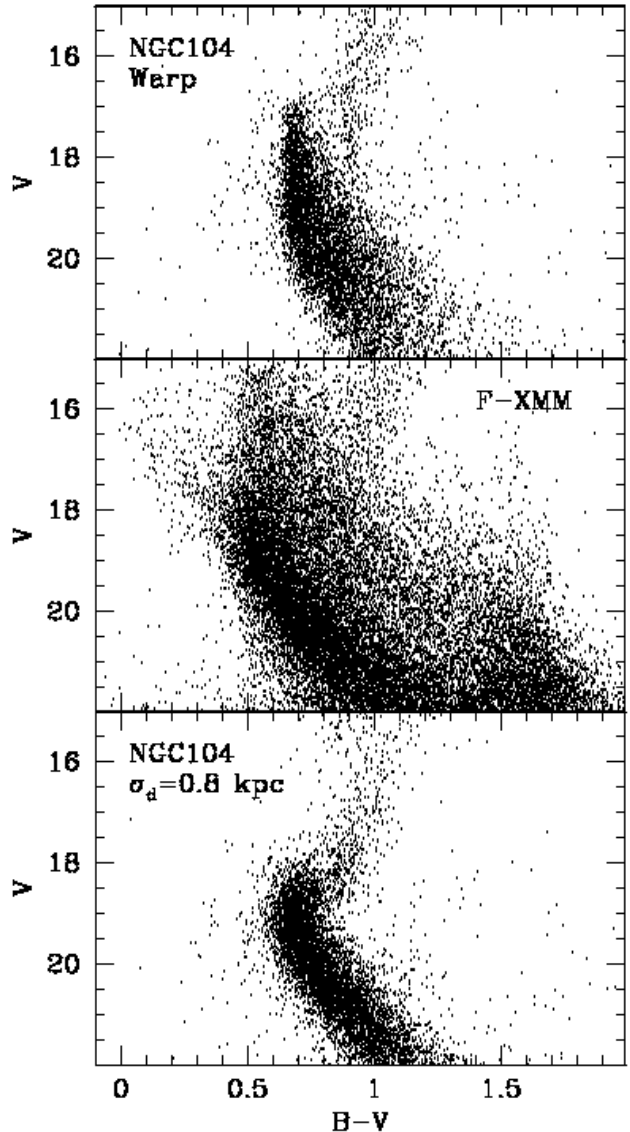} 
\caption{Synthetic (upper and lower panels) and observed (middle panel) Colour
Magnitude Diagrams toward $(l;b)=(244.2\degr;-8.2\degr)$. Upper panel: 
CMD of $\sim 20000$ stars randomly extracted from the photometry of the 
cluster 47 Tucanae by \citet{47tuc}, as it would appear if the considered stars
were distributed according to the distance distribution predicted by the 
Lopez-Corredoira at al. Warp model in the above direction. Middle panel:
observed CMD toward $(l;b)=(244.2\degr;-8.2\degr)$ (XMM Field, from Pap-II).
Lower panel: CMD of $\sim 20000$ stars randomly extracted from the photometry of the 
cluster 47 Tucanae by \citet{47tuc}, as it would appear if the considered stars
were distributed according to a Gaussian distribution having mean $D_{\sun}=7.2$
kpc and standard deviation $\sigma=0.8$ kpc, i.e. the line of sight distance
distribution of CMa as estimated by \citet{delgado-cmd}.}
\end{figure*}

Despite the large differences shown in Fig.~6, it may still be argued that 
the L02 Warp model still predicts some South-North overdensity 
at the position of CMa. Are the observed and
predicted distributions {\em locally} similar, at this position? To test this
possibility we recur to the optical CMD of the XMM field (F-XMM, at $l=244.2\degr$ and
$b=-8.2\degr$) we presented in Pap-II, that is reproduced in the middle panel of
Fig.~7. To have an idea of how the CMD of an ensemble of old
stars distributed according to the L02 model would appear we proceeded as follows: (1) we
take as template population the photometry of $\sim 130000$ stars of the
globular cluster 47 Tucanae presented by \citet{47tuc}, (2) we extracted, at
random, 20000 stars from this sample and we corrected their magnitude for
extinction and for distance 
\cite[assuming E(B-V)=0.04 and (m-M)$_0$=13.29 for the custer, from][]{f99}, (3)
we assigned to each star a distance randomly extracted from the distance
distribution predicted by the L02 model toward $(l,b)=(244.2\degr, -8.2 \degr)$, and
(4) we corrected for the corresponding distance moduli and for the reddening of
the F-XMM field (E(B-V)=0.14, see Pap-II). 

The {\em synthetic} CMD thus obtained is shown in the upper panel of Fig.~7. Its
morphology is completely different from that of the observed diagram. Stars
around the Turn Off (TO) point of the cluster form a vertical sequence from
$V\sim 17.0$ to $V\sim 20.0$ that abruptly bends to the red for $V>20.0$ in the
way typical of CMDs of Galactic fields that sample populations at widely different
distances along the line of sight. This result is similar to what we
obtained in Pap-II while comparing the F-XMM CMD with the predictions of the
warped and flared Galactic model by \citet{r03}. Note that $\sim $30\% of
the stars plotted in the upper panel of Fig.~7 have associated distances that
place them in the regime in which we adopt an extrapolation of the L02 model.
However, repeating the experiment while forcing the selection of stars with
$R_{GC}\le 13.0$ leads to a CMD having a morphology essentially
indistinguishable from that shown here. Hence, also in this case the adopted
extrapolation of the L02 model has a negligible effect on our results.
In contrast, the F-XMM CMD shows a
clear and narrow Main Sequence that bends continuously from the TO (around
$V\sim 18.5$) to the limit of the survey, clearly resembling a bound stellar
system. This similarity is made clearer by the comparison with the CMD shown
in the lower panel of Fig.~7. This synthetic diagram has been obtained repeating
the steps 1, 2 and 4 described above but assigning to stars distances extracted
from a Gaussian distribution having a mean $D_{\sun}=7.2$ kpc (i.e., the
distance of CMa as derived in Pap-III) and $\sigma=0.8$ kpc, as derived by
\citet{delgado-cmd} for CMa. Except for the foreground field population that is,
obviously, present in the F-XMM and absent in the higher latitude 47 Tuc field,
the two MS features are strikingly similar. This test confirms that the observed CMa
overdensity in not predicted by current models of the Galactic Warp, neither on
global or local scales. While all the considered models predict that the
Galactic Warp should provide some overdensity at the position of CMa, in good
agreement with the results shown in Fig.~3, none of
them is able to reproduce anything remotely similar to the extremely compact,
cored structure we observe in Canis Major.

It should be recalled that the Galaxy models including a warped and flared outer
Disc, as those considered here and in Pap-II and Pap-III, are not meant to
provide a perfect representation of the Galaxy, in particular for what concerns
the outer parts of the Galactic Disc where the available observations are scanty
and uncertain over large ranges of longitude. Hence, the fact that these models
are unable to reproduce the observed structure of CMa is not an ultimate proof
that Canis Major cannot be a substructure of the Galactic Disc. On the other
hand, this is a proof that (a) Canis Major is a truly unexpected structure,
since no similarly spatially confined and dense structure is included in
existing warped and flared models of Galactic Disc and, (b) the
CMa structure is fairly different from the general expectations of an
overdensity due to the Galactic Warp, since it is a compact elongated and cored
structure instead of a large wedge-shaped lobe, as shown in Fig.~6, and, 
above all, it appears at a significantly different position.

\subsubsection{The Argo structure}

While  the present  paper was  near completion  a preprint  was posted
\cite[][hereafter  RP05]{argus} studying the  distribution of  2MASS M
giants  at  low galactic  latitudes  ($|b|\le  20\degr$). Adopting  an
extinction grid with a  finer resolution compared to previous analyses
of  M-giants \citep[][Pap-I]{majewski,crane},  and using  a starcounts
model to  subtract the  smooth Galactic components,  they are  able to
follow overdensities  down to  lower absolute latitudes  ($|b|$) than,
for instance,  in Pap-I.   They find that,  in the $220\degr  \le l\le
320\degr$  region,  all the  existing overdensities  are
enclosed in  the range of Heliocentric distances  $6.0\kpc < D_{\sun}<
20\kpc$.  They  confirm the  presence of the  CMa overdensity  in this
range but they claim that it is part of a larger and denser structure,
centred  around $l\simeq 290\degr$,  that they  interpret as  a large
dwarf galaxy  which they name {\em  Argo}. They suggest  that Argo and
CMa can  be erroneously seen as  separated entities only  because of a
plume  of  extinction  located  at  $l\simeq  265\degr$  extending  to
$b\simeq  -12\degr$   (see  Fig.~2 and Fig.~8, below)  that   introduces  
an  artificial discontinuity in the  surface density. 

In Fig.~8 we present the
South-North subtracted surface density map of RC stars in the 
range $200\degr  \le l\le 320\degr$, obtained by counting stars (in the distance
range 5.0 kpc \ltsima $D_{\sun}$ \ltsima 13.0 kpc)
within $4\degr \times 4\degr$ boxes spaced by $2\degr$ both 
in longitude and latitude. In this case also stars with $|b|<5.0\degr$ have 
been included in the sample, for a better comparison with RP05.
The  main body of the claimed Argo structure is clearly traced, in surface
density, also by RC stars. In Fig.~8 it appears as a strong, elongated 
overdensity with a broad maximum around $l\sim 290\degr-310\degr$, with CMa
placed at its low longitude edge (compare  with Fig.~1  and Fig.~3  of RP05). 

The maps displayed in Fig.~3a,b (as well as Fig.~5, 6, and 7, above) show that
{\em there is no spatially confined 3-D structure} corresponding to this strong 
{\em surface} overdensity. Integrating over the range $6.0\kpc < D_{\sun}<
20\kpc$ along the $l\sim 280\degr-300\degr$ direction implies summing up the 
contribution of all the stars encountered along the line of sight that crosses 
the maximum of the overdensity produced by the Galactc warp. 
In any case, while in terms of
surface density - integrated over 14 kpc along the line of sight - CMa is a much
weaker structure compared to Argo (see Fig.~8), it corresponds to a much 
more {\em spatially confined} structure: 
CMa spans the range 5 kpc  $\la D_{sun} \la$ 12 kpc, with a line-of-sight
Half Width at Half Maximum of just $\sim 2.0$ kpc (see Pap-II,
\citet{delgado-cmd} and Sect.~4.1, below).  The
major  axis of  CMa  is approximately  oriented  along the  path of  a
nearly-circular orbit around the centre  of the Galaxy (see Fig.~5 and
Fig.~11, below), consistent with the N-body simulation presented in Pap-IV
(see Sect.~4.2, below). On the other hand, the Argo structure corresponds to a
part of the large-scale southern overdensity that spans more than 15 kpc in
heliocentric distance, toward $l\sim 290\degr$ and it does not appear as a
coherent, cored structure as CMa (see, Fig.~3).

\begin{figure}
\includegraphics[width=84mm]{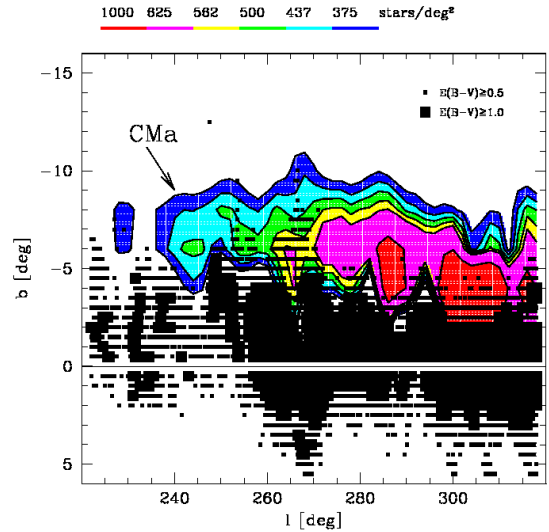} 
\caption{Subtracted map (South-North) in terms of surface density in the sky for
RC stars in the range $11.5\le K\le 14.0$. The surface density has been computed
on a grid of $4\degr \times 4\degr$ pixels spaced by $2\degr$ both in latitude
and longitude. In this case stars with $|b|<5.0\degr$ have also been included in
the sample. The position of CMa is indicated by a labelled arrow.
Small squares mark positions
where the reddening is  $0.5\le E(B-V)< 1.0$, large squares correspond
to reddening  in excess of  1.0 mag.}
\end{figure}

The above arguments are clearly displayed in Fig.~9, where we consider the
surface density of RC stars confined to the very narrow range of Galactocentric
distances including most of the main body of CMa 
(12.0 kpc $\le R_{GC}\le$ 14.0 kpc). 
The subtracted surface density map in the upper
panel of Fig.~9 has been obtained in the same way as Fig.~8, but is expressed
in terms of fractional density excess to allow a comparison with the L02 model.
It is quite evident that at this spatial location the only strong overdensity is
CMa, while Argo has completely disappeared. On the other hand the fractional
excess due to the Warp (integrated over $5\degr \le |b|\le 15\degr$, Fig.~9,
lower panel), 
as predicted by the L02 model, reaches its maximum at the longitude 
corresponding to the maximum density of Argo ($l\sim 290\degr - 300\degr$).

There is also a quantitative self-consistency problem to consider. RP05 derive
(from M  giants) an estimate of the total luminosity of $L=3.7 - 15\times 10^6 
~L_{\sun}$ for Argo.  Limiting ourselves to the stars contributing to the map 
of Fig.~8 and within  $265\degr \le l\le 320\degr$ we count an excess of more
than $1.8\times 10^6$ RC stars.  With the assumptions stated in  Sect.~4.3,
below, we obtain  a total luminosity of $L\sim 1.2\times  10^9$,  orders  of
magnitude larger  than the  R05  estimate.   Note that  our  estimate should 
be considered as a  lower limit since (a) the contribution  of CMa is not
included, while RP05 consider it as a part of Argo, and (b) we limited the
calculation to the approximate distance range $5\kpc\le D_{sun}\le 13\kpc$,
while R05 integrate over $6.0\kpc < D_{\sun} < 20\kpc$. Hence the beautiful
consistency among the luminosity estimates obtained from M-giants,  RC stars 
and  MS  stars achieved  for  CMa (see Sect.~4.3, below)   is completely lost
in the case  of Argo.  This strongly suggests that Argo is  not a  coherent 
stellar system,  but  is instead  just a  surface density  structure that 
originates  from the  mix  of different  (and unrelated)  populations that 
happen to  lie  along the  same line  of sight.

\begin{figure}
\includegraphics[width=84mm]{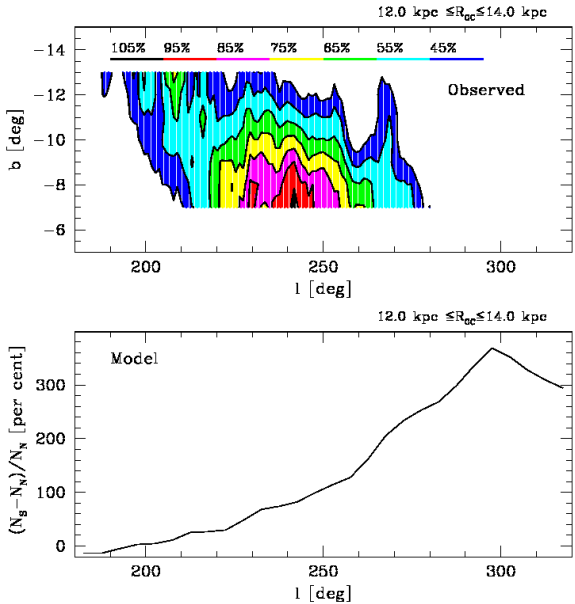} 
\caption{
Upper panel: 
subtracted map (South-North) in terms of fractional excess of surface
density in the sky for RC stars in the range of Galactocentric
distances 12.0 kpc $\le R_{GC}\le$ 14.0 kpc, i.e. the range enclosing the
densest part of the CMa structure.
Lower panel: fractional density excess as a function of Galactic longitude in 
the same range of Galactocentric distances, as predicted by the
L\'opez-Corredoira et al. Warp model. Data for $|b|<5.0\degr$ are not included here.}
\end{figure}

RP05 discard the warp hypothesis for the nature of Argo claiming
that the  shape of  the structure is  not ''warp-like''.   However, in
their view, {\em all} the  southern overdensities in the $220\degr \le
l\le  320\degr$  region  are  due  to  Argo  and  CMa.   The  implicit
conclusion is that the Galactic  Warp (and flare) does not produce any
observable  South-North asymmetry  in this  region, at  odds  with the
conclusions  of  \citet{lopez,vig}.   The  smooth  nature  of  the  global
South-North  asymmetry  shown  to  be  present  in  this  region,  its
Galactic-scale  dimensions, and the  fact that  a rescaled  version of
Northern  density map  is  a  good model  for  it (Fig.~5),  strongly
indicate   that  the   surface  density   structure   centred  around
$l=290\degr$ is  not a dwarf galaxy  remnant (Argo) but  the effect of
the Warp (and flare) asymmetries of the Galactic disc.

Turning back our attention to CMa, there is an interesting point that can be
appreciated from the comparison of Figs.~9 and 10. While the CMa structure is
present in both maps, its appearance is quite different. Most of the difference
is just apparent and is due to the density scale adopted in Fig.~8, that is
best suited to provide a good view of ''Argo'' and compress the whole structure
of CMa to just three levels of density. However part of the difference is real,
and is due to the much larger degree of foreground/background contamination
affecting the map of Fig.~8, which includes stars at any Galactocentric
distance. In particular, the center of the structure (i.e., the region of
maximum density) is shifted from $l\sim 242\degr$ in the distance limited map
(Fig.~9, that should be closer to the actual physical structure of the
system) to $l\sim 244\degr$ in the all-distances map of Fig.~8, due to
contamination by likely unrelated stars at various distances along the line of
sight. This kind of effect should be always taken into account in studying
the structure of CMa, since effects due to the {\em integration along the line of sight} 
can significantly change the overall view of the system.

\subsection{The nature of the Canis Major overdensity}

In our view there are only two possible explanations for the observational
evidence presented above:

\begin{enumerate}

\item Canis Major is a strong and unexpected substructure of the outer Galactic
Disc. This would imply that the Disc hosts very dense and compact substructures
of {\em old stars} (the age should be larger than 1-2 Gyr to have RC stars and
should be larger than 4-5 Gyr according the optical CMDs, see Pap-II and
\citet{delgado-cmd}). Such substructure should be very different, in nature,
from usual spiral arms, for instance, that host populations dominated 
by very young stars.
 A local distorsion of the outer Disc, coupled with a large
scale warp, may also possibly reproduce the observed distribution of
overdensities in the Southern Galactic Hemisphere.

\item Canis Major is the remnant of a disrupting dwarf galaxy, as suggested in
Pap-I, II, III,  and IV. This interpretation naturally fits all the
observational evidence collected up to now, with particular emphasis on the
size and density of the structure and to the kinematics of its stars (see
Pap-III, Pap-IV and \citet{delgado-pm}).

\end{enumerate}

For the reasons described above and in the previous papers of this series, our
preference goes to the second hypothesis, that appears to provide a much easier
explanation of the whole observational scenario. In the following we will take
this hypothesis as true and, on this basis, we will derive some interesting
physical characteristics of CMa as a stellar system.

\section{Physical parameters of Canis Major}

\subsection{Distance}

In Pap-III we obtained a distance profile of CMa using M-giants as distance
indicators. Here we can repeat that analysis using the more reliable distance
scale provided by RC stars. In the upper panel of Fig.~10 we show the distance
distribution of the RC stars in our sample in the Northern (blue) and Southern
(red) hemispheres, in a wide region toward the centre of CMa. 
The southern profile shows an
overall excess at any distance, with respect to the northern one, plus an
additional bump peaked around $D_{\sun}\simeq 7.5$ kpc. We rescaled the northern
distribution by the same factor as in Fig.~5 (e.g., $\times 1.2$) and subtract
it from the southern profile. The result is shown in the lower panel of Fig.~10:
the South-North excess appear to be confined within $D_{\sun}\simeq 5$ kpc 
and  $D_{\sun}\sim 12$ kpc. The peak of the profile is at 
$D_{\sun}=7.2 \pm 1.0$ kpc and the Half Width at Half Maximum is $HWHM\simeq
2.0$ kpc, in excellent agreement with the results of Pap-I, Pap-III and
\citet{delgado-cmd}. 
There are a few interesting considerations that can
be deduced by inspecting Fig.~10: 

\begin{itemize}

\item The  distance profile of  CMa as obtained  from its RC  stars is
strikingly  similar to  that derived  from M-giants  in  Pap-III. Even
leaving  aside  the  perfect  coincidence  of the  peaks  of  the  two
distributions, that may  be ascribed to a mere  coincidence within the
uncertainties associated  with the two  distance scales, it  remains a
fact that the two independent  tracers give essentially {\em the same}
line-of-sight profile of the CMa system.

\item In our various analyses  we have obtained distance estimates for
CMa from four different  and independent distance indicators. In order
of  accuracy: from  the tip  of the  Red Giant  Branch in  the  I band
($D_{\sun}\simeq  7.2$ kpc,  Pap-III), from  the Red  Clump in  K band
($D_{\sun}\simeq  7.2$ kpc,  present analysis),  from  the photometric
parallax of M giants in  the K band ($D_{\sun}\simeq 7.1$ kpc, Pap-I),
and from Main Sequence fitting  in the B and V bands ($D_{\odot}\simeq
7.6 -  8.7$, Pap-II).  All  of these independent  methods consistently
provide  a  distance  of  the  centre  of  CMa  within  $7.0\kpc$  and
$9.0\kpc$. Hence, the  average distance of CMa should  be now regarded
as a very reliable piece  of information about this system.  Note that
if $D_{\sun}\simeq  7.2\kpc$ is assumed, the age  estimate obtained in
Pap-II would  shift toward older ages  (i.e.  6 Gyr$\la$  age $\la 10$
Gyr, instead of 4 Gyr$\la$ age $\la 10$ Gyr).

\end{itemize}

\begin{figure}
\includegraphics[width=84mm]{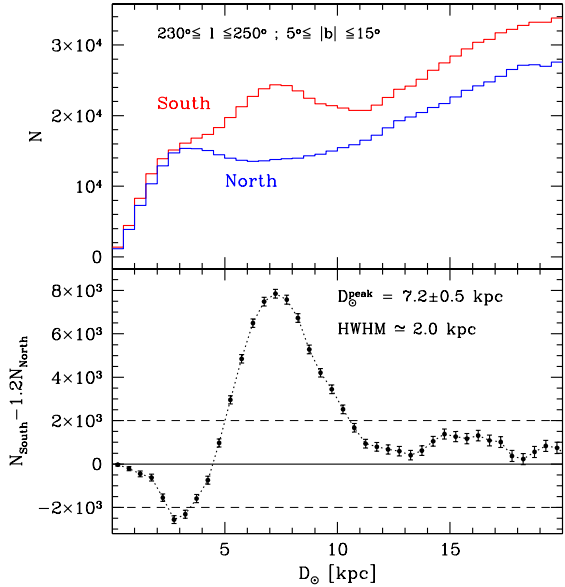} 
\caption{Distance profile of the Canis Major overdensity. Upper panel: distance
distribution of RC stars in the range $230\degr \le l\le 250\degr$ and 
$5\degr \le |b|\le 15\degr$, in the Southern (Red) and Northern (blue) Galactic
Hemisphere. Lower panel: residual of the subtraction of the rescaled Northern
profile (by a factor $\times 1.2$) from the Southern profile. 
The long-dashed lines approximately encloses the range of variation of the 
flat part of the residual in response to variations of the scaling factor by
$\pm$ 10 per cent.}
\end{figure}

\subsubsection{The spatial orientation of CMa}

The subtracted density maps of Figs.~3, 5 and 6 make evident that the physical
orientation of the elongated body of CMa implies a  heliocentric 
distance-longitude gradient: the heliocentric distance of CMa grows with
Galactic longitude. Using the same technique as above we find that the distance
of the main body of CMa varies from $D_{\sun}\simeq 6.3$ kpc at $l\simeq
225\degr$, to $D_{\sun}\simeq 7.2$ kpc at $l\simeq 240\degr$, 
to $D_{\sun}\simeq 9.3$ kpc at $l\simeq 265\degr$.

The effect is clearly depicted in Fig.~11, where the contours of the
main body of CMa (in red, taken from Fig.~5b) are superposed on the
results of the  N-body simulation of the disruption  of the CMa galaxy
presented in  Pap-IV.  The  simulation is produced  by a  dwarf galaxy
model of mass $5\times 10^8 \msun$ that is accreted onto the Milky Way
during  $\sim 3$  Gyr (see  Pap-IV for  details) and  it was  built to
reproduce the current mean position and 3-D velocity of CMa as well as
the  {\em distance--radial  velocity} gradient  discovered  in Pap-IV.
Hence the {\em distance--longitude} behaviour of the remnant is a {\em
prediction} of the model, not an {\em a priori} requirement. Therefore
the excellent qualitative  and quantitative  agreement with  the {\em
observed} and {\em predicted} spatial  orientation of the main body of
CMa  may be  considered  as a  success  of the  dwarf-galaxy-accretion
scenario for the  nature of CMa.  In other words, the structure and spatial
orientation of CMa as obtained from RC stars is  fully consistent
with the  model of  a disrupting dwarf  galaxy that fits  the observed
position and kinematics of CMa.

\begin{figure}
\includegraphics[width=84mm]{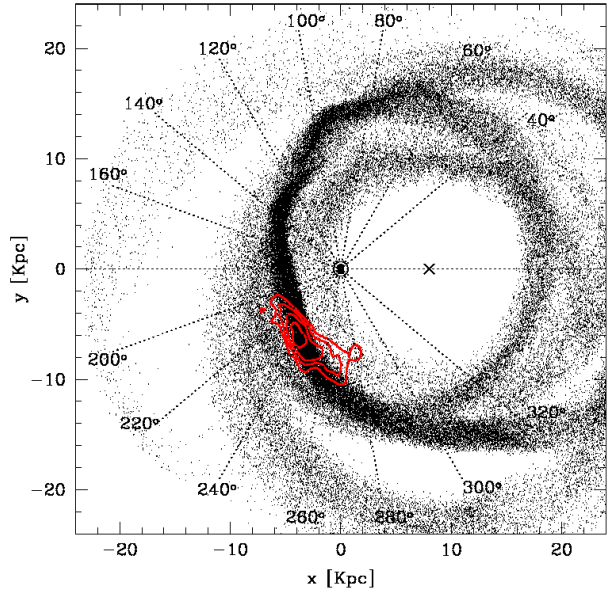} 
\caption{The excess-density contours of CMa taken from Fig.~5b (down to
$16\sigma$, in red) are compared with  the N-body model  of the
disruption  of  the  CMa  dwarf  galaxy  presented  in  Pap-IV  (small
dots).}
\end{figure}

\subsection{Luminosity and surface brightness}

The Evolutionary  Flux relation \citep{rb86, rf88,  alvio} states that
the total number of stars in a given evolutionary phase $j$ ($N_j$) is
proportional   to   the  total   luminosity   of   a  stellar   system
($L_T$).  Considering  a  Simple   Stellar  Population  (SSP,  i.e.  a
population of stars with the same age and chemical composition):
\begin{equation}
N_j = B(t) L_T t_j
\end{equation}
where $t_j$  is the duration of  the evolutionary phase  in years, and
$B(t)$ is  the specific  evolutionary flux, i.e.  the number  of stars
entering or leaving any post Main Sequence evolutionary phase per year
per solar luminosity  as a function of time ($t$, i.e.  the age of the
SSP).

We  can use Eqn.  4 to  obtain a  distance-independent estimate  of the
total luminosity of CMa from the  RC star counts. With some algebra we
obtain:
\begin{equation}
M_V = -2.5 \Big[ \log[N_j] - \log[{L_T\over L_V}] - \log[t_j] -
  \log[B(t)] \Big] + M_{V\sun}
\end{equation}
where $M_V$  is the  absolute integrated magnitude  in the  V-band and
$M_{V\sun}$ is  the absolute V magnitude  of the Sun.  For the present
case we adopt $B(t)=1.71\times 10^{-11}$ and ${L_T\over L_V}=1.5$ from
the SSP  models by \citet{claudia, claudia2}  with $[M/H]=-0.33$, age$=
6.0$ Gyr and Salpeter Initial Mass Function
\footnote{see  http://www.mpe.mpg.de/~maraston/SSP}.  $M_{V\sun}=4.83$
is also  adopted according to  \citet{claudia}, while the  duration of
the  core  He-burning phase  ($t_{HB}=0.87\times  10^{8}$  yrs) for  a
population with  age$\simeq 6.0$ Gyr and $[M/H]=-0.35$  has been taken
from \citet{p04}\footnote{see http://www.te.astro.it/BASTI/}.

In  the same  way  we  can use  Eqn.  5 by  \citet{alvio}  to obtain  a
distance-dependent estimate of the  surface brightness near the centre
of CMa ($\mu_{V,0}$):
\begin{equation}
\mu_{V,0} = -2.5 \log(n_j) + 2.5 \log(FoV) + (m-M)_0 + Norm
\end{equation}
where $n_j$ is the number of stars in the $j$ evolutionary phase within the
considered Field of View (FoV, in $arcsec^2$, for example), and the term
$Norm$ encompasses all the theoretical assumptions:
$$Norm = 2.5 \log[{L_T\over L_V}] + 2.5 \log[t_j] +2.5 \log[B(t)] + M_{V\sun}$$
We assume  the distance modulus $(m-M)_0=14.3$, as  derived in Pap-III
and in the present contribution.

To estimate  the observables required by Eqn.~5 and
Eqn.~6 ($N_{HB}$ and  $n_{HB}$) we proceeded as follows. We selected from our
sample the RC stars of both hemispheres within $200\degr \le  l \le  280\degr$,
$15.0\degr  \le |b| \le  5.0\degr$ and 10.0 kpc $\le R_{GC}\le$ 16.0 kpc 
(the adopted selection is displayed in Fig.~12). 
Then we subtracted the total number of
Northern RC stars - rescaled by the usual 1.2 factor - from the total number of
Southern RC stars in this region. We obtain an excess of $\simeq 1.1\times 10^5$
RC stars that can be associated to CMa, corresponding to a total luminosity of
$L_T\simeq 7.4\times 10^7 L_{\sun}$. The integrated absolute magnitude is 
$M_V=-14.4\pm 0.8$, where the uncertainty includes a $\pm 10$ per cent 
variation in the $\times 1.2$ scaling factor adopted to take into account the 
contribution of the large-scale overdensity (Warp) to the detected excess. 
This estimate is  in excellent
agreement  with  those  obtained  in   Pap-I  from  M  giants  and  by
\citet{delgado-cmd} from optical  photometry ($M_V=-14.5\pm 0.1$).   
We estimate that the inclusion or exclusion from the sample of 
the stars within the $|b|<5\degr$ region may change  the value of the 
V-band  integrated magnitude by $\sim \pm 0.5$ mag.  

To estimate the surface brightness we computed the South-$1.2\times$ North 
excess of RC stars within a $2\degr \times 2\degr$ box centred on 
$(l,b)=(240\degr,-6\degr)$ and on $(l,b)=(244\degr,-6\degr)$. We chose two
different positions to provide estimates of the surface brightness for the
centre of the structure as obtained from a distance-selected sample (as in
Fig.~9) and from the straight surface density distribution, as in Fig.~8. We
obtain a density of $\simeq 314$ RC stars ${\rm deg^{-2}}$ at $l=240\degr$,
corresponding to $\mu_V=24.0 \pm 0.6$ mag/arcsec$^2$, and $\simeq 370$ RC stars
${\rm deg^{-2}}$ at  $l=244\degr$, corresponding to $\mu_V=23.8 \pm 0.6$
mag/arcsec$^2$. These values are in reasonable agreement with the results of
\citet{delgado-cmd} obtained from data  at $(l,b)=(240\degr,-8\degr)$ ($\mu_{V,0}=23.3
\pm 0.1$ mag/arcsec$^2$).

Note  that $B(t)$,  at  fixed $[M/H]=-0.33$,  varies from  $1.66\times
10^{-11}$  at $t=1.0$ Gyr  to $1.75\times  10^{-11}$ at  $t=13.0$ Gyr,
and, for fixed  age $= 6.0$ Gyr, it  ranges from $1.66\times 10^{-11}$
at  $[M/H]=-1.35$  to  $1.98\times  10^{-11}$ at  $[M/H]=+0.35$.   The
variation of  $t_{HB}$ over the same  range of age  and metallicity is
just of  a few per cent  \citep{p04}. Hence the  particular choices of
the above parameters  do not seriously affect our  final estimates. 
Other factors, not completely accounted for, as the uncertainty in the 
exact position of the center of CMa, the overall extent of the system, and 
the adopted SSP hypothesis, are expected to be the dominant contributors 
to the uncertainty of our  estimates of the integrated magnitude and of 
the central surface brightness of the system.

CMa  turns out to  have a  total luminosity  and a  surface brightness
quite similar to those of the Sgr dSph, as already noted in Pap-I, but
the above  results suggest that it  may be slightly  more luminous and
with a brighter  $\mu_V,0$, in agreement with the  $M_V ~vs. ~\mu_V,0$
relation  of  dwarf   spheroidal  galaxies  \cite[see][and  references
therein]{delgado-cmd}. 

It  is interesting to note  the striking effect
of  the relative  distance on  the  number densities  for two  stellar
systems  of  similar  central surface  brightness.   \citet{lorenzoHB}
measured  a density of  RC stars  of $\simeq  2100$ deg$^{-2}$  in the
central  degree of  the Sgr  dSph, while  here we  observe  just $314$
deg$^{-2}$ at $l=240\degr$,  yet  the  surface   brightness  of  the  
two  systems  is similar.  
This may  largely account  for the  great difficulty  in the
identification of the  RC in optical CMDs from  off-centred fields of
$0.25$  deg$^{2}$  or  less  as  those  presented  in  Pap-II  and  in
\citet{delgado-cmd}. This argument will  be developed further
in Sect.  4.4, below.

\begin{figure}
\includegraphics[width=84mm]{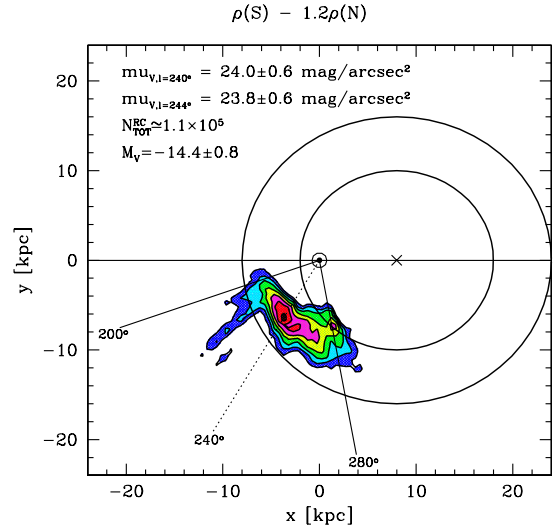} 
\caption{Subtracted density map taken from Fig.~5a (Northern Hemisphere
densities rescaled by a factor $\times 1.2$). The continuous circles with
radius $R_{GC}=10.0$ kpc and $R_{GC}=16.0$ kpc, and the continuous lines
indicating the directions $l=200\degr$ and $l=280\degr$ enclose the region that
we have considered for the computation of the total number and the
central surface density of RC stars attributable to Canis Major. These numbers
have been used to compute the integrated absolute magnitude and the surface
brightness (toward two different directions) of the system, that are reported in
the figure.}
\end{figure}

\subsection{How should a nearby galaxy appear?}

CMa is  by far the nearest  dwarf galaxy known, a  factor $\simeq 3.6$
closer   than   the  previous   ``record   holder'',   the  Sgr   dSph
\citep{lorenzotip}.   This  is  the  main  reason why  it  appears  so
unusual, at first glance: a  structure covering tens of degrees on the
sky  and with as  few as  $\sim 300-400$  RC stars  in its  central square
degree.  This is very far from our idea of the typical dwarf satellite
of the Milky Way, that has been acquired by studying much more distant
examples.  To show that  there is  nothing unusual  in CMa  except its
distance we undertook  the exercise of seeing how  the Sgr dSph galaxy
would appear  if it were placed  at the same distance  and position as
CMa.  In the following we  will adopt the structural parameters of Sgr
(core radius $r_c$, limiting  radius $r_t$, central surface brightness
and  axis ratio)  as derived  by \citet{majewski},  the  distance from
\citet{lorenzotip} and the surface  density of Horizontal Branch stars
from \citet{lorenzoHB}.

Fig.~13 shows the core and the limiting contours of Sgr once placed at
the distance and  position of CMa. The upper panel  of Fig.~13 has the
same  longitude scale  as  Fig.~9, to allow a direct  comparison. 
The core of
the  ``mock'' Sgr  shows  a major  axis  of more  than $\sim  26\degr$
($r_c=13.3\degr$) and the overall  appearance is strikingly similar to
CMa. Note  that the limiting contour extends  well beyond $l=200\degr$
on the low longitude side, and beyond $l=300\degr$ on the other 
($r_t=64.5\degr$). Note also that the outskirts of the ``mock'' Sgr
galaxy extends into the Northern Galactic Hemisphere, 
up to $b\sim +5/10\degr$.

From Eqn.~6 the following relation  can be derived between the surface
brightness  (at any  chosen  wavelength $\lambda$)  $\mu_{\lambda,1}$,
$\mu_{\lambda,2}$ and  the number of stars  per given area  in a given
evolutionary  phase ($n_1,n_2$)  for  two stellar  systems located  at
distances $d_1$ and $d_2$:
\begin{equation}
\mu_{\lambda,1}-\mu_{\lambda,2} = -2.5\log(n_1/n_2) + 5\log(d_1/d_2)
\end{equation}

The second term  on the right hand side of  Eqn.~7 displays the effect
of  distance on  the number  surface density  of a  system of  a given
surface brightness\footnote{It has to be  noted that the effect of the
physical size of a system on  the spread of sequences on a CMD weakens
with increasing  distance. This is  an additional factor  that favours
the easier recognition of  distant stellar structures compared to more
nearby systems.}.  To rescale the  number surface density of  Sgr from
$d_2=26.3\kpc$      to      $d_1=7.2\kpc$      we     must      assume
$\mu_{\lambda,1}-\mu_{\lambda,2}=0.0$.   In the  present case  we find
that  the observed  surface  number  density of  Sgr  stars should  be
reduced by a factor 0.075, if  the galaxy is placed at the distance of
CMa.   Hence, in  this  case, we  would  observe $\sim  150$ RC  stars
deg$^{-2}$, 10 RR Lyrae deg$^{-2}$ and 12 Blue Horizontal Branch stars
deg$^{-2}$ in the centre of the Sgr galaxy.  This effect fully
accounts for the difficulty  of detecting clearly-defined sequences of
evolved stars in small fields (smaller than 1 deg$^2$) even within the
main body  of CMa,  as well as  for the low yield in searches  of rare
stellar species  \cite[as RR Lyrae stars  in a system  dominated by an
intermediate-age and metal rich population, see][]{kinman,mateu}.

In conclusion, the apparent size of  CMa and the number density of its
stellar populations are fully compatible  with those of a dwarf galaxy
similar to Sgr and located at $7.2$ kpc from us.

\begin{figure}
\includegraphics[width=84mm]{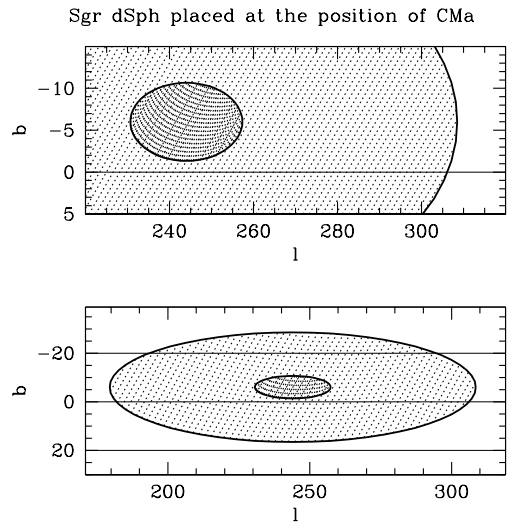} 
\caption{The  appearance of  the Sgr  dSph if  it were  placed  at the
distance and position  of CMa. The inner ellipse  encloses the core of
the galaxy (densely-shaded region),  the outer ellipse is the limiting
contour     \citep[sparsely    shaded    region;     Sgr    parameters
from][]{majewski}.  The  upper panel has  the same longitude  scale as
Fig.~8,  to allow a direct comparison with the map shown there.}
\end{figure}

\subsection{Old open clusters in CMa}

The  possible association  of  some  old open  clusters  with CMa  was
already suggested in  Pap-I \cite[see also Pap-II;][]{crane,frinch}. A
thorough analysis of these clusters is clearly beyond the scope of the
present contribution.  However,  the use of the Red  Clump as a tracer
may provide the unique opportunity  of placing a few clusters onto the
same distance  scale as  the field population  of CMa.  In  Table~1 we
list a  number of open clusters  selected to be older  than $\sim 0.8$
Gyr  and  to  lie  in  the  surroundings  of  the  main  body  of  CMa
\cite[according  to the WEBDA  database,][]{mermi}. In  particular the
first group  presented in  Tab.~1 is composed  of three  clusters that
directly  project onto  the main  body  of CMa  (as seen in Fig.~8 and 10, for
instance), and  for which  it has  been  possible to
estimate  the K  magnitude of  the  Red Clump  from 2MASS  photometry.
Hence  the distance  estimates for  To~2,  AM~2 and  Haf~11 have  been
derived  in  exactly the  same  way  and  under the  same  assumptions
(luminosity  of  the  standard  candle  and reddening)  used  for  the
analysis  of the  field population  performed above.   These estimates
strongly indicate  that To~2, AM~2 and Haf~11  are  physically located
within the  main body  of CMa (see  Pap-II for further  discussion and
references). Note that, while AM~2 and Haf~11 lack any radial velocity
estimate, To~2  has $V_r\simeq +114$  km/s, fully consistent  with the
systemic velocity of CMa (Pap III, Pap IV).

\begin{table*}
 \centering
 \begin{minipage}{140mm}
  \caption{Old open clusters around the main body of CMa.}
  \begin{tabular}{@{}lcccccc@{}}
  \hline
   Name & l$\degr$ & b$\degr$ & E(B-V) &$K^{RC}_0$ & $(m-M)_0$ & $D_{sun}$ [kpc] \\
  \hline
 Tombaugh~2 & 232.83 & -6.88 & 0.29 & $12.95\pm 0.10$ & $14.45\pm 0.22$ & $7.76\pm 0.8$\\ 
 Haffner~11 & 242.39 & -3.54 & 0.57 & $12.20\pm 0.20$ & $13.70\pm 0.28$ & $5.49\pm 0.8$\\
Arp-Madore~2& 248.12 & -5.88 & 0.51 & $13.65\pm 0.10$ & $15.15\pm 0.22$ & $10.71\pm 1.1$\\ 
\hline
 Berekeley~36 & 227.38&  -0.59 &&&&  6.1\\
 NGC~2243     & 239.48& -18.01 &&&&  4.4\\ 
 Melotte~66   & 259.56& -14.24 &&&&  4.3\\ 
 Ruprecht~75  & 276.79&  -4.48 &&&&  4.3\\
 CC06                   & 238.48&  -4.28      &&&&	 4.5\\
 Van den Bergh-Hagen~66 & 276.00&  -3.01      &&&&	 7.0\\
 Saurer~2               & 257.99&  -1.01      &&&&	 4.8\\
\hline
 \footnote{All the
  reported data, with the exception of those relative to the RC distance of
  To~2, AM~2 and Haf~11, are drawn from the WEBDA database \citep{mermi},
  see http://obswww.unige.ch/webda/. Data for CC06 are from \citep{ivanov}.
  First group: clusters within $230\degr \le l\le 260\degr$, 
  $-12\degr \le b\le 3\degr$
  and 4.0 kpc $\le D_{sun}\le $11.0 kpc, with age $\ge 1$ Gyr.
  Second group: clusters within $220\degr \le l\le 280\degr$,
   $-20\degr \le b\le 0\degr$
  and 4.0 kpc $\le D_{sun}\le $11.0 kpc, with age $\ga 0.8$ Gyr.}
  \end{tabular}
\end{minipage}
\end{table*}


\section{Summary and Conclusions}

We have performed  a differential analysis of the  distribution of the
Southern Galactic hemisphere excess  of colour-selected Red Clump stars
extracted  from 2MASS in  the region  $40\degr\le l\le  320\degr$ and
$5\degr \le |b|\le 15\degr$.  Our main aim was to trace the structure
of the  main body of the  newly discovered Canis  Major stellar system
(Pap-I)  and of  the stellar  component of  the southern  part  of the
Galactic Warp (Pap-II,  Pap-III). 

We have obtained detailed maps of the
South-North and North-South overdensities in the Galactic Disc. These
maps show that the third and fourth Galactic quadrants ($180\degr\le l\le
360\degr$) are dominated by a smooth large-scale South-North asymmetry that
extends all over the considered half of the Disc and over a large range of
heliocentric distances. The above characteristics, the comparisons with the
large-scale North-South asymmetries observed in the first and second quadrants
($0\degr\le l\le 180\degr$) and the comparison with the warped and flared
Galactic model of L02 strongly suggest that such a wide South-North overdensity is
due to an asymmetry of Galactic scale, in particular the southern lobe of the
Warp of the Galactic Disc. 

The Canis Major structure, as recognized in Pap-I and Pap-2, appears - in the
South-North density maps - as the strongest {\em spatial} overdensity of the
whole Galactic Disc, either in terms of number density or of statistical
significance.
Simply subtracting a
rescaled version of the Northern hemisphere density map (by a factor 1.2) from
the Southern hemisphere density map, the large scale asymmetry attributable to
the Galactic Warp is completely removed and Canis Major
remains as  the only significant overdensity, with a peak at more than 25
$\sigma$. The structure appears elongated along the tangential direction,
extending from $l \simeq 200\degr$ to $l \simeq 280\degr$. It has a nearly
elliptical shape and it is strongly spatially confined, with a HWHM along the
line of sight at its center of $\simeq 2.0$ kpc \cite[in excellent agreement 
with the results of Pap-I, Pap-III, and][]{delgado-cmd}. All the maps show that
Canis Major has a roundish core, located at $D_{\sun}=7.2\pm 1.0$ kpc, toward
$l=241.7\degr$, at a Galactocentric distance of $R_{GC}=13.1\pm 1.0$ kpc.

The L02 model of the Galactic Warp is unable to predict the existence of
Canis Major as a structure of Galactic origin, both on the large and on the
local scale. The same has been shown to be true also for the \citet{yusi} and
the \citet{r03} models, in Pap-III and in Pap-II, respectively (see also
Fig.~1, above). Within the
framework set by the observational evidence listed above, Canis Major 
appears as a compact and confined stellar system  that is  ``superposed'' on 
the  Warp structure (see Fig.~3, for example).

We  also   investigate  the  claim   made  in  the  recent   study  by
\citet{argus}, that Canis Major is  an external field of a larger Argo
structure, which would be the  true centre of the dwarf galaxy remnant
accreted onto the Milky Way. The distribution of red clump stars along
the line of sight does  not support this conclusion, however. Unlike the
situation in the Canis Major region, we find no significant spatially-confined
structure in the line of  sight through Argo. This line of sight, instead,
crosses the region of the disc were the large-scale overdensity associated with
the Galactic Warp reaches its maximum.  
We conclude that  Argo is most likely not related to Canis Major, and is 
probably an asymmetry of Galactic origin.

The above observational scenario does not provide the ultimate word on the
nature of CMa. Given our poor knowledge of the outermost region of the Galactic
Disc and the obvious limitations in current models of the Disc itself, the
possibility that CMa is a substructure of the Galactic Disc cannot be excluded
yet. However, in this case, we should admit the presence of an unexpected
large and very dense substructure and/or local distorsion 
in the outer Disc, hosting essentially  {\em old} stars, at odds with, 
for instance, usual spiral arms.

In our view, the evidence collected here, in Pap-I, Pap-II and by
\citet{delgado-cmd}, together with the kinematics of Canis Major stars as
described in Pap-III, Pap-IV and the possible links with the Monoceros Ring
\cite[Pap-I, Pap-IV,][]{delgado-mod,delgado-pm,dana,connb} are more naturally
explained by the hypothesis that Canis Major is the remnant of the disrupting
dwarf galaxy in a nearly-circular and nearly-planar orbit about the centre of 
the Galaxy. 
Accepting this hypothesis as the one that best fits all the available
data we have derived several physical characteristics of CMa as a stellar
system, that are summarised in Table~2. In particular we draw the attention of
the reader on the following specific results of the present analysis:

\begin{itemize}

\item RC  stars provide a quantitative  description of CMa  that is in
full  agreement with  those  obtained from  other independent  tracers
\cite[M   giants   and   MS   stars,  see   Pap-I,   Pap-II,   Pap-III
and][]{delgado-cmd}.      In      particular     we     obtained     a
(distance-independent) estimate  of the  integrated V magnitude  and a
(distance-dependent)  estimate of  the central  surface  brightness in
good  agreement with those  obtained by  \citet{delgado-cmd} from
deep  optical photometry  of MS  stars.  We  also obtained  a distance
profile  (along  the  line  of  sight)  of  CMa  that  is  essentially
indistinguishable  from that obtained  in Pap-III  from M  giants. The
high degree  of self-consistency  achieved among the  various analyses
--- using different tracers --- may be considered also as a validation
of our criterion of selection  of RC stars.   
We confirm,  with  independent tracers  and an  independent
distance  scale with  respect  to  Pap-I and  Pap-III,  that the  mean
distance of CMa is $D_{\sun}=7.2 \pm 1.0$ kpc, around $l=240\degr$.

\item We have  detected a clear relation between  the mean distance to
CMa and  Galactic longitude  within the main  body. The
derived spatial  orientation of the  system is in good  agreement with
the  predictions of the  N-body simulation  presented in  Pap-IV, that
models CMa as a dwarf galaxy being accreted in a planar orbit onto the
disc of the Milky Way.

\item Using the  same dataset (2MASS), the same  tracer (RC stars) and
the same distance scale as in the analysis summarised above, we showed
that the  old open clusters AM~2,  To~2 and Haf~11  are located 
(in space) within the main body of Canis Major.

\end{itemize}

\begin{table}
 \centering
 \begin{minipage}{89mm}
  \caption{Fundamental parameters of the main body of the Canis Major system.}
  \begin{tabular}{@{}lc@{}}
  \hline
$\langle D_{sun}\rangle$ & $~~~~~~~~~~~7.2\pm 1.0$ kpc\\
$l_0$\footnote{As derived from Fig.~8. Adopting the distance selected 
sample of Fig. 10, $l_0=241.7\degr \pm 1.0\degr$ is obtained (see Sect. 3.2.1).} & $~~~~~~~~~~~244.0\degr \pm 2.0\degr$\\
$b_0$\footnote{As derived from Fig.~8.} & $~~~~~~~~~~~-6.0\degr \pm 1.0\degr$\\
$\langle R_{GC}\rangle$ & $~~~~~~~~~~~ 13.1 \pm 1.0$ kpc\\
x & $ ~~~~~~~~~~~3.7 \pm 0.6$ kpc\\
y & $~~~~~~~~~~~ -6.4 \pm 0.6$ kpc\\
z & $ ~~~~~~~~~~~\simeq -0.8 $ kpc\\
$M_V$ & $~~~~~~~~~~~-14.4 \pm 0.9$ mag\\
$\mu_{V,240\degr}$ & $~~~~~~~~~~~24.0 \pm 0.6$ mag/arcsec$^2$\\
$\mu_{V,244\degr}$ & $~~~~~~~~~~~23.8 \pm 0.6$ mag/arcsec$^2$\\
$HWHM_{l.o.s.}$ & $~~~~~~~~~~~\simeq 2.0$ kpc\\
\hline
\end{tabular}
\end{minipage}
\end{table}


In summary,  the present study fully (and  independently) confirms the
results  presented   in  previous  papers   of  this  series   and  in
\citet{delgado-cmd}: Canis Major has  the size, the luminosity and the
kinematics  typical of  a large  dwarf  galaxy, seen  during the  last
stages  of its  disruption within  the tidal  field of  the  Milky Way
\cite[see also][]{boniUVES}. While a Galactic origin for CMa cannot 
be definitely ruled out with the present data, the dwarf galaxy hypothesis
appears as best suited to fit the overall observational scenario.

While further  details emerge --- as, for
example,  the distance-longitude  gradient ---  we begin  to  obtain a
clearer view of this challenging system.  Future studies should try to
overcome the  several observational  challenges posed by  this nearby,
low  latitude object,  to unveil  the details  of its  stellar content
(age,  chemical composition)  and to  clarify its  possible connection
with the Monoceros Ring.

\section*{Acknowledgments}

The  financial  support  of   INAF  and  MIUR  is  acknowledged.   
M.B. is grateful to the ULP/Observatoire de Strasbourg for the kind hospitality
during a period in which a significant part of the analysis presented in this
paper was performed. The Referee, M. L\'opez-Corredoira, is
acknowledged for his very helpful comments that led us to a much deeper
analysis of the scientific problems considered here.
This publication makes  use of  data products from  the Two Micron  All Sky
Survey, which  is a joint  project of the University  of Massachusetts
and the  Infrared Processing and  Analysis Center/California Institute
of  Technology,   funded  by   the  National  Aeronautics   and  Space
Administration and the National Science Foundation.  This research has
made use of NASA's Astrophysics Data System Abstract Service.
The assistance of P. Montegriffo in the development of the software 
required for the present analysis is also acknowledged.

\appendix
\section{Contamination by dwarfs}

To study the effect of the contamination by dwarfs in our color-selected 
sample of candidate RC stars on subtracted density maps, we recur to the
Galactic model by \citet{r03}. This model includes the effect of the Galactic
Warp and of the disc flaring, and takes into account the off-plane position of
the Sun. The main advantages of using the R03 model for our purpose can be
summarized as follows: 

\begin{itemize}

\item we can obtain (K,J-K) CMDs of the synthetic samples, therefore we can 
apply exactly the same selections and analysis adopted for the observed 
sample;

\item synthetic stars from the R03 models are individually classified into
different species with the flag $cl$, hence the discrimination of dwarfs from
giants is straightforward. In particular stars having $cl=3$ are
giants, those with $cl=4$ are subgiants and those with $cl=5$ are dwarfs.
Since the synthetic CMDs reveal that both $cl=4$ and $cl=5$ stars can fall into
our selection color-window, from now on we will consider all the stars of class
4 or 5 as {\em dwarfs}.

\end{itemize}

The obvious disadvantage is that the result will not reflect the {\em real}
Galaxy but just a model of it. However, in the present context, the unavoidable
approximations inherent to a Galactic model should not represent a major concern.
In the following we will mainly focus on the magnitude range $12.0\le K\le
14.0$, that is the range enclosing essentially all the stars that contribute
to the main structures we discuss in the present paper, and CMa in particular. 

To allow the possibility of a direct comparison with observed CMDs (a very
useful sanity check) and to limit the dimensions of the synthetic catalogues we
choose to work on $30$ deg$^2$ windows in the latitude range 
$6\degr\le |b|\le 11\degr$, i.e. exactly the same as the observed sample whose
CMD is shown in Fig.~1. To study the dwarf contamination toward different
directions of interest, we selected three windows at $l=180\degr \pm 3\degr$, 
$l=241\degr \pm 3\degr$ (i.e., the same as Fig.~1), and $l=280\degr \pm 3\degr$.
We extracted from the model the corresponding samples in the Southern and
Northern Galactic hemispheres, at the above longitudes, for a total of six
catalogues. The extraction was performed using the {\em large field} option,
that takes into account the variations of the population across the considered
fields \footnote{See the web form of the R03 model and the explanations provided
therein, http://bison.obs-besancon.fr/modele/.}. 
To reproduce as well as possible the real case we let the model 
include the effects of interstellar extinction according to the default
assumptions included in the web form.

From the observed dataset of Fig.~1 we fitted the following relations for the
photometric errors in  K(J) as a function of - not extinction-corrected - K(J)
magnitude:

\begin{equation}
\epsilon K = 0.0236-0.00047K+3.480\times10^{-8}exp(K) ~~mag~~
\end{equation} 

\begin{equation}
\epsilon J = 0.0169+0.00024J+5.840\times10^{-9}exp(J) ~~mag~~.
\end{equation} 
  
To simulate realistically the effect of photometric errors on synthetic stars
we computed $\epsilon K$ and $\epsilon J$ for each of them with the above
equations, we randomly extracted a Gaussian deviate ($\delta$, positive or
negative, in $\sigma$  units) for each of them, and then we added the resulting
synthetic errors $\delta_K\times\epsilon K$ and $\delta_J\times\epsilon J$ to
their K and J magnitudes, respectively. After that, we corrected the synthetic
stars for extinction, using the $A_V$ value associated by the model to each
star. In the following J and K stand for extinction-corrected magnitudes, as in
the rest of the paper.

\begin{figure}
\includegraphics[width=84mm]{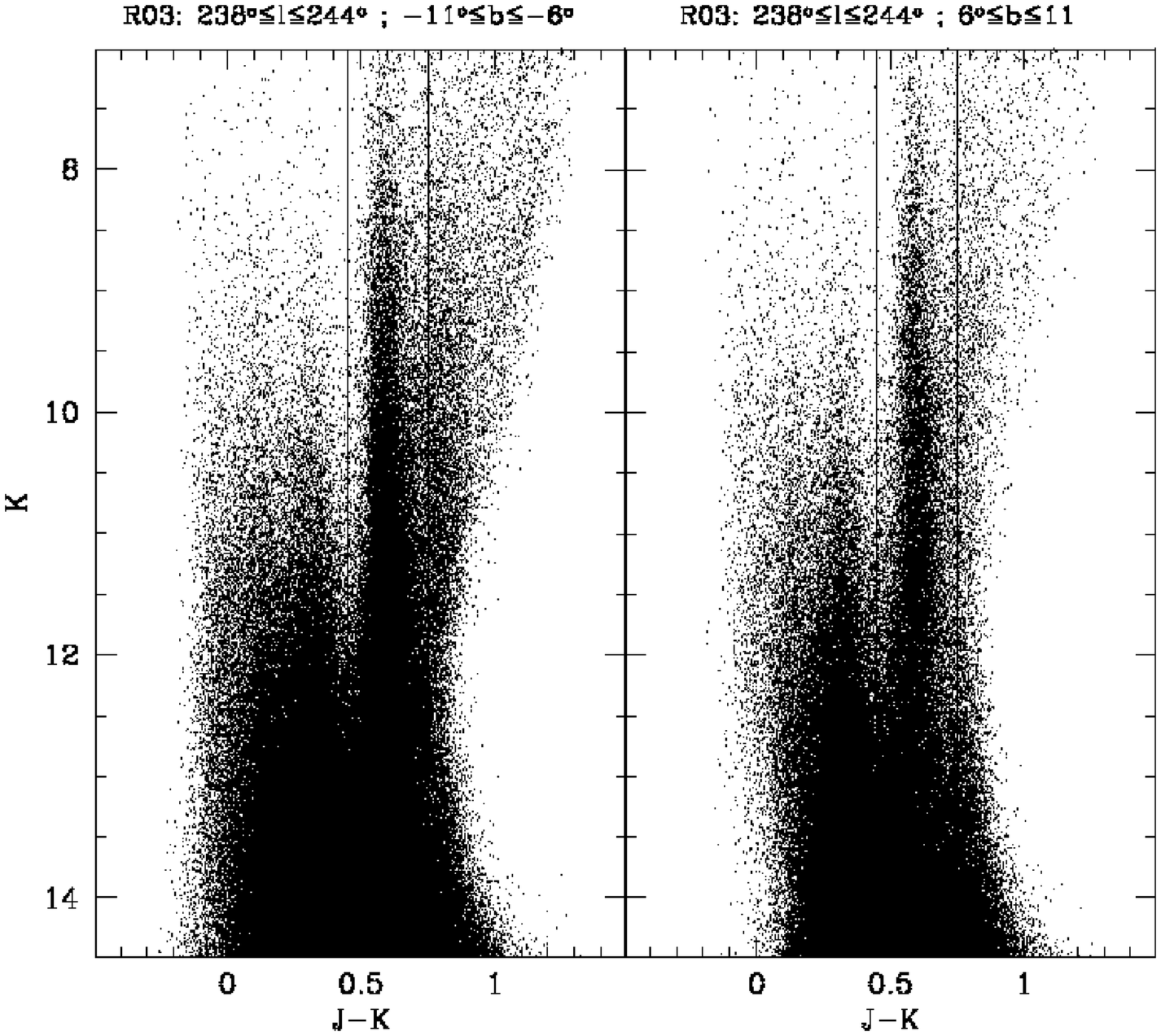} 
\caption{Example of synthetic CMDs extracted from the R03 model. The symbols 
and the scale are the same as in Fig.~1. The continuous line encloses the
color-window $0.45\le J-K\le 0.70$.}
\end{figure}

The above procedure should provide synthetic samples as similar as possible to
the observed ones, including the effect of photometric errors. The comparison
of the final synthetic CMDs for the $l=241\degr$ window (Southern and Northern
hemisphere) shown in Fig.~A1 with the observed CMD of Fig.~1 demonstrates that
this objective has been achieved: the overall similarity between observed and 
synthetic CMDs is striking. In Fig.~A1 we report also the adopted color
window we use to select RC stars in the observed sample.  A shift of $0.04$ mag
in J-K has been applied to all the synthetic samples to center the vertical
plume of RC stars within the selection window, as it is in the observed case.
This small mismatch in color is likely due to a non-perfect match between the
theoretical star tracks used by the R03 model and real stars (quite a usual
occurrence) as well as to tiny differences in the adopted extinction laws. In
any case the following results are essentially unaffected by the small color
shift applied. In the following we will consider only syntethic stars falling
in our color-selection window.  Hence, when we speak of {\em dwarfs} we mean
stars of class 4 or 5 (R03) having $0.45 \le J-K\le 0.70$, and for {\em giants}
we mean stars of class 3  having $0.45 \le J-K\le 0.70$, that is, the stars
that we would have selected as RC stars in our analysis.

Before proceeding with our set of tests, Fig.~A1 deserves a few comments. It is
quite clear that there is a strong asymmetry between the Southern and Northern
samples: the Southern sample contains many more stars and the difference is
particularly evident in the RC plume. This is the effect of the Warp included in
the model. However the comparison of the LFs of the observed (Fig.~1,
left panel) and of the synthetic (Fig.~A1, left panel) Southern samples shown in
the right panel of Fig.~1 demonstrates that (a) the R03 model provides an
excellent overall representation of the observed distribution of stars in the
considered color window (except for the total normalization, which appears too
large by a factor $\sim 3$), and (b) it still fails to reproduce the strong bump
in the observed LF around $K\simeq 13.0$, i.e. the putative RC of CMa.
This comparison is fully consistent with the coesistence of the Galactic Warp
{\em and} an additional stellar system, CMa. Finally, the excellent agreement
between the syntetic and observed LFs outside the $K=13.0\pm 0.5$ range, seems
to confirm that our samples are essentially complete down to $K\sim 14.5$, as
stated in \citet{cutri}.

\begin{figure}
\includegraphics[width=84mm]{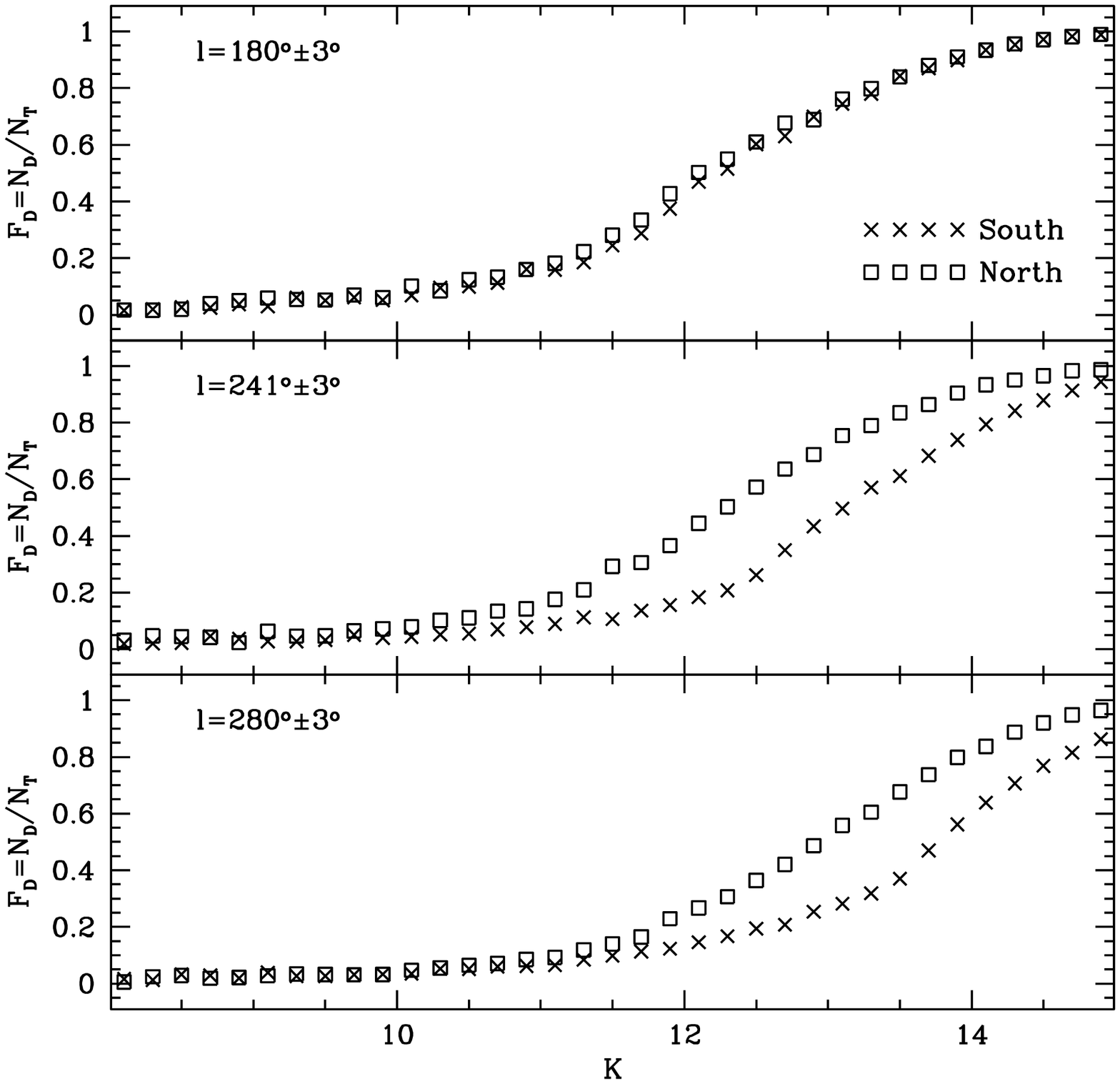} 
\caption{Fraction of dwarf stars in the total number of color-selected RC stars
as a function of K magnitude. The three panels show $F_D$ for three different
directions, in the Northern (open squares) and Southern ($\times$ symbols)
hemispheres.}
\end{figure}

In Fig.~A2 the fraction of dwarfs falling into our color-selection as a function
of K magnitude ($F_D$) is shown for the Northern (open squares) and 
Southern (crosses) samples for the three selected directions. 
There are several interesting conclusions that can be drawn from this plot:

\begin{itemize}

\item First of all, since the line of nodes of the warp included in the R03
model lies exactly toward the $l=180\degr$ direction,  any South-North
difference in $F_D$ and/or in the detected South-North asymmetries in the
$l=180\degr$ sample must be due to the  effect of the $\sim 15$ pc displacement
of the Sun to the North of the Galactic Plane. The upper panel of Fig.~A2 shows
that  difference in $F_D$ is very small, less than 5 per cent in any bin  for
$K\ge 12.0$ and  $\simeq$ 1 per cent in the whole $12.0\le K\le 14.0$ range.
Moreover, the South-North difference in giants at $l=180\degr$ is 5 per cent of
that at $l=241\degr$ and 2 per cent of that at $l=280\degr$. Finally the
South-North difference in the {\em total} number of color selected RC stars at
$l=180\degr$ is 12 per cent of that  at $l=241\degr$ and 6 per cent of that at
$l=280\degr$. The conclusion is that the off-plane position of the Sun  has
very little effect both on the contamination by dwarfs and on the amplitude of
observed South-North asymmetries, for the applications considered here. Hence
our decision to neglect it in the analysis described in this paper appears
to be fully justified.

\item In the range most relevant for the present study ($12.0\le K \le 14.0$)
$F_D$ is quite large, reaching the 80 per cent level at $K=14.0$, in some cases.

\item The overall degree of contamination is quite similar independent of
the considered directions, from the Anticenter to $l=280\degr$.

\item There are significant differences in the Northern and Southern $F_D$
toward $l=241\degr$ and $l=280\degr$, up to $20-30$ per cent in some bins.
However $F_D$ is {\em always higher in the Northern hemisphere than in the
Southern one}.  It appears that, according to the R03 model, N-S differences
in the contamination by dwarfs are working {\em against} the detection of
real overdensities in the Southern hemisphere.

\end{itemize}

The latter result may appear puzzling at a first glance, but it can be
easily explained by looking into the synthetic samples.  It turns out
that it is due to the presence of a significant South-North asymmetry
at large Galactocentric distances (i.e.  the Warp) that is much larger
in giants than in dwarfs, in the magnitude ranges of interest.  For
example, in the range $12.0\le K\le 14.0$, at $l=241\degr$ the
South-North ratio of dwarf density is
$\rho_D(S)/\rho_D(N)=29096/24371=1.19$ while for giants
$\rho_G(S)/\rho_G(N)=28684/7572=3.79$.  Therefore, even if there are
more dwarfs in the South than in the North, the imbalance in giants is
much larger and the final $F_D$ is consequently larger in the North
($F_D=$76 per cent) than in the South ($F_D=$50 per cent).  At
$l=280\degr$, $\rho_D(S)/\rho_D(N)=35427/31333=1.13$, while
$\rho_G(S)/\rho_G(N)=70305/22067=3.18$, in the same range of
magnitudes.  This implies that in presence of a distant South-North
asymmetry (as the Warp, as in the R03 model we are presently
considering, or CMa, or both), the observed imbalance in dwarfs is
small in this magnitude range, such that the contribution by dwarfs
essentially disappears when Southern and Northern counts are
subtracted.

The above results drive us to the two key questions we have to answer
to really check the reliability of the results of the present paper:

\begin{enumerate}

\item What is the contribution of dwarf contamination to 
     {\em subtracted densities}?

\item Is our observable, the subtracted density of color selected RC
stars $\rho(S)-\rho(N)$, a good tracer of the real overdensity of
giants? In other words, is it a good estimator of the quantity we are
interested in ($\rho_G(S)-\rho_G(N)$)?

\end{enumerate}

\begin{figure}
\includegraphics[width=84mm]{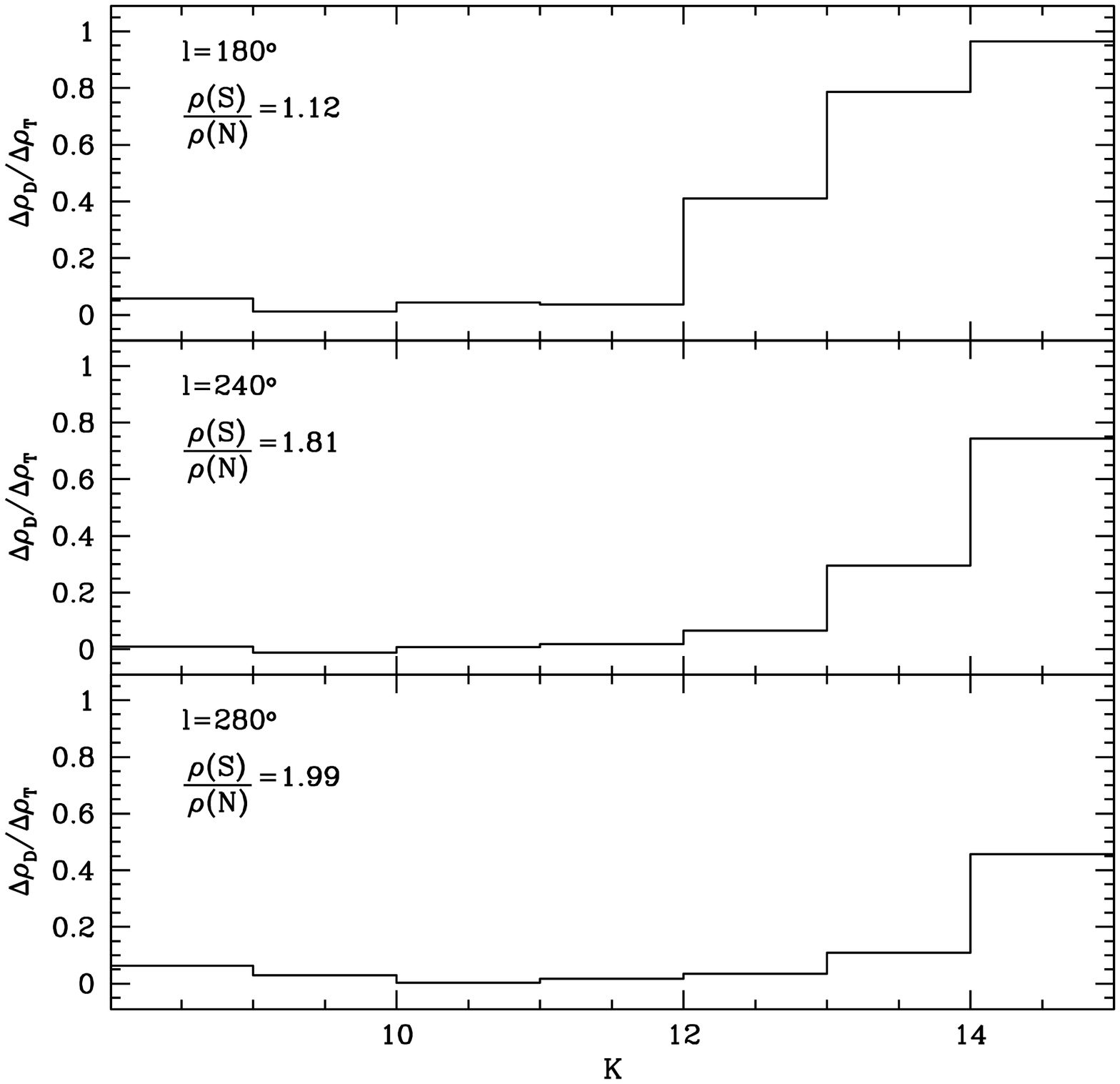} 
\caption{Contribution of the South-North imbalance in dwarf stars to the total
South-North density difference as a function of magnitude, for three different
directions. $\rho(S)/\rho(N)$ is the ratio of color-selected RC stars in the
range $12.0\le K\le 14.0$, and roughly describes the degree of total South-North
imbalance in each give direction.}
\end{figure}

Fig.~A3 answers to the first question. There we plot the fraction of the total
S-N density difference that is due to dwarfs 

$${{\Delta \rho_D}\over{\Delta \rho_T}}= 
{{\rho_D(S)-\rho_D(N)}\over{\rho(S)-\rho(N)}}$$ 

as a function of magnitude, for the three considered directions. The subscript T
stands for Total$=$Dwarfs$+$Giants.  The ratio of the total Southern-to-Northern
density is also reported within each panel. This plot shows that in absence of
a real South-North asymmetry ($l=180\degr$, $\rho(S)/rho(N)\sim 1$) the
contribution is quite significant ($>$ 40per cent) for $K>12.0$. On the other
hand, toward directions that cross the Warp asymmetry ($l=241\degr, 280\degr$,
$\rho(S)/rho(N)>1.8$) the dwarf contribution remains lower than 30 per cent
everywhere, for $K\le 14.0$. Hence, even in a regime of {\em strong}
contamination by dwarfs, the effects on South-North subtraction are quite
limited for $K\le 14.0$ and in the presence of an overdensity like the Warp and/or
CMa. On the other hand, our subtracted maps may be significantly affected (or
even dominated, in some cases) by dwarf  contamination around $l=180\degr$ and
for $K>14.0$. It is very interesting to note that in those regions we see a
low-significance and smooth South-North residual (Fig.~3) that completely
disappears when we subtract by the rescaled Northern density map (Fig.~5). On
the other hand all the structured overdensities we deal with in this paper lie
in the region where dwarfs provide a minor contribution to the subtracted
densities ($l>200\degr$ and $K\le 14.0$). Within this context, there is another
interesting point worth considering. Since the contribution of dwarfs to
our subtracted densities grows continuously with magnitude even for $K\le
14.0$, if our maps were seriously affected  by dwarf contamination we would
expect to see a corresponding growth of density with heliocentric distance. On
the contrary, in the direction of CMa we see  {\em both a growth and a decrease
of the density} (see Figs.~3 and 5). The distance profile of Fig.~11 shows that
toward $l\simeq 240\degr$ the  subtracted density reaches a maximum at
$D_{\sun}=7.2$ kpc, corresponding to $K=12.8$, and then essentially falls to
zero at $D_{\sun}\simeq 10.0$ kpc, corresponding to $K=13.5$, a completely
different profile with respect to what is expected from a distribution of Galactic
dwarf stars.  This occurrence strongly suggests that dwarfs cannot have a
major role in determining the observed distribution of the CMa overdensity. 
The same argument is valid also when considering the trend of contamination of
subtracted densities as a function of Galactic longitude. Since the degree of
contamination grows toward $l=180\degr$, if dwarfs were dominant contributors to
the subtracted density we should see the maximum density toward $l=180\degr$.
On the contrary, the density in that direction is low and smooth,
while it is much higher and more strongly peaked toward $l\simeq 240\degr$.

\begin{figure}
\includegraphics[width=84mm]{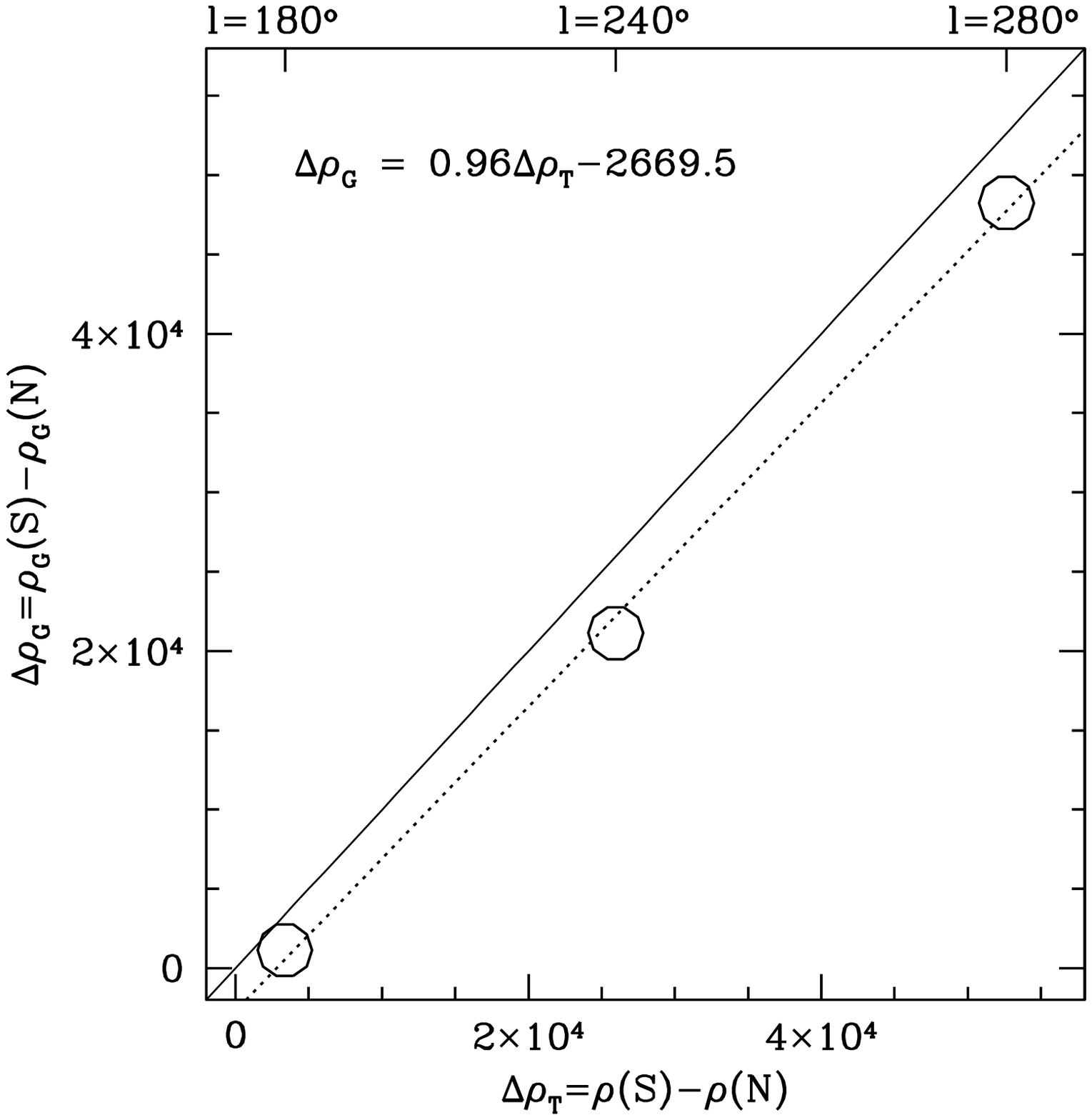} 
\caption{Correlation   between  the   total   subtracted  density   of
color-selected   RC  stars   and  the   real  subtracted   density  of
{giants}. The continuous line passes  through the origin and has slope
1.0, while the dotted line is the best linear fit to the data (see the
equation  in the  upper  left  corner of  the  plot).  The  directions
corresponding to the three plotted points are indicated in the upper x
axis.}
\end{figure}

Fig.~A4 tries to answer question (ii), above.  This is simply a plot
of the S-N density difference in the total number of stars selected in
color as RC candidates with our technique (i.e., our observable
$\Delta \rho_T$), versus the actual difference in the S-N density of
giants (i.e.  the quantity we want to estimate, $\Delta \rho_G$), in
the range $12.0\le K\le 14.0$, for the three directions we are
considering.  The correlation between $\Delta \rho_T$ and $\Delta
\rho_G$ is {\em very strong and essentially linear}.  Moreover the
cofficient of the linear best-fit model is near unity, hence {\em our
observable appears as an excellent estimator of the true South-North
density difference in giants}, at least in the presence of an
asymmetry like the warp embedded in the R03 model.

In summary, based on the above experiments with the R03 Galactic model, 
we can conclude that:

\begin{itemize}

\item The neglection of the displacement of the Sun with respect to the
Galactic plane has very little effect on our analysis, and can be safely
ignored.

\item While our samples of color-selected RC stars may be heavily
contaminated by dwarfs, the effects of the contamination are much
lower on the subtracted-density maps. The contribution of dwarfs to
the ``observed'' $\Delta \rho_T$ is lower than 30 per cent for $K\le
14.0$ toward $l=240\degr$ and $l=280\degr$, and much lower than this
for $K\le 13.0$. Note that these numbers take into account only known
Galactic components, the presence of an extra asymmetry in the
considered range of distances (i.e. CMa, see Fig.~1) would further
reduce the r\^ole of dwarfs.

\item The observable we adopt ($\Delta \rho_T$) appears to trace very
well the true South-North imbalance of giants under the realistic
conditions provided by the R03 model.

\end{itemize}

It is conceivable that the CMa overdensity we observe using
color-selected putative RC stars may be in fact produced by an
unexpected asymmetry in dwarfs.  However the dwarfs contributing to
our counts in the range $12.0 \le K\le 14.0$ have heliocentric
distances 0.0 kpc $\le D_{\sun}\le $ 4.5 kpc. Their distance
distribution shows a {\em strong} peak at $D_{\sun}\simeq 1.0$, the
mean is $\langle D_{\sun}\rangle = 1.0$ kpc and the standard deviation
is $\sigma=0.7$ kpc, and 89 per cent have $D_{\sun}\le 2.0$ kpc. Hence
a strong and {\em very compact} clump of dwarfs\footnote{Actually it
must be significantly more compact than the distribution of RC stars
we attribute to CMa, since the selected dwarfs would span a much
larger range of absolute magnitudes, with respect to RC stars.} would
have to be present in the Southern hemisphere, at $\sim 1.0$ kpc from
the Sun toward $l=240\degr$, a feature that is obviously {\em not
present} in the R03 model. As far as we know there is no report in the
literature of such a local overdensity. Finally, it should be recalled
that the position and size of the RC overdensity studied here matches
very well those derived for CMa with other tracers (M giants and MS
stars) that are much less affected by contamination from spurious
sources.

\label{lastpage}

\end{document}